\documentclass[rmp,aps,twocolumn,floatfix,superscriptaddress,eqsecnum,nofootinbib,showpacs,longbibliography]{revtex4}
\usepackage{amsmath,amssymb,amsfonts}
\usepackage{latexsym}
\usepackage{graphicx}
\usepackage{dcolumn}
\usepackage{bm}
\usepackage{psfrag}
\usepackage{subfigure}
\usepackage{tabularx}
\usepackage{natbib}
\usepackage{hyperref}
\hypersetup{
   linktoc=all,     %set to all if you want both sections and subsections linked
}
\usepackage{color}

\def\be{\begin{equation}}       \def\ee{\end{equation}}
\def\bea{\begin{eqnarray}}      \def\eea{\end{eqnarray}}
\def\ba{\begin{array} }
\def\ea{\end{array} }
\def\bnum{\begin{enumerate} }
\def\enum{\end{enumerate}}

\def\=>{\Rightarrow}
\def\>{\rightarrow}

\def\eye2{Fathbb{I}}

\def\LSCO{La$_{2-x}$Sr$_x$CuO$_4$}
\def\LBCO{La$_{2-x}$Ba$_x$CuO$_4$}
\def\YBCO{YBa$_2$Cu$_3$O$_{6+y}$}

\def\BSCCO{Bi$_2$Sr$_2$CaCu$_2$O$_{8+\delta}$}
\def\oxychloride{Ca$_{2-x}$Na$_x$CuO$_2$Cl$_2$}
\def\LNSCO{La$_{1.6-x}$Nd$_{0.4}$Sr$_x$CuO$_{4}$}

\def\C60{A$_x$C$_{60}$}

\def\HgCu3{HgCa$_2$Cu$_3$O$_{8+y}$}
\def\HgCu4{HgBa$_2$Ca$_3$Cu$_4$O$_{10+y}$}
\def\TlCu{Tl$_2$Ba$_2$CuO$_{6+\delta}$}
\def\TlCu3{Tl$_2$Ba$_2$Ca$_2$Cu$_3$O$_{10+y}$}
\def\TlCu4{Tl$_2$Ba$_2$Ca$_3$Cu$_4$O$_{12+y}$}

\def\BiCu3{Bi$_2$Sr$_2$Ca$_{2}$Cu$_3$O$_y$}

\def\8LSCO{La$_{1.88}$Sr$_{.12}$CuO$_4$}
\def\110LNSCO{La$_{1.5}$Nd$_{0.4}$Sr$_{0.1}$CuO$_{4}$}
\def\stage4LCO{La$_{2}$CuO$_{4+\delta}$}
\def\Y248{YBa$_2$Cu$_4$O$_8$}

\def\NbSe2{NbSe$_2$}
\def\TaSe2{TaSe$_2$}
\def\TiSe2{TiSe$_2$}
\def\NaCoOH2O{Na$_{0.3}$CoO$_{2y}$H$_2$O}
\def\MgB2{MgB${}_2$}

\def\URu2Si2{URu$_2$Si$_2$}
\def\hty{high-temperature superconductivity}
\def\hts{high-temperature superconductors}

\begin{document}

\title{Theory of Intertwined Orders in High Temperature Superconductors}
\author{ Eduardo Fradkin}
\affiliation{Department of Physics \&\ Institute for Condensed Matter Theory, University of Illinois at Urbana-Champaign, Urbana,
Illinois 61801-3080, USA}
\author{Steven A. Kivelson}
\affiliation{Department of Physics, Stanford University, Stanford, California 94305-4060,
USA}
\author{ John M. Tranquada}
\affiliation{Condensed Matter Physics \&\ Materials Science Department, Brookhaven National Laboratory, Upton, NY 11973-5000, USA}
\date{\today}
\begin{abstract}
The electronic phase diagrams of many highly correlated systems, and in particular the cuprate high temperature superconductors, are complex, with many different phases  appearing with similar---sometimes identical---ordering temperatures even as material properties, such as a dopant concentration, are varied over wide ranges.  This complexity is sometimes referred to as ``competing orders.'' However, since the relation is  intimate, and can even lead to the existence of new phases of matter such as the putative ``pair-density-wave,'' 
  the general relation is better thought of in terms of  ``intertwined orders.''  We selectively analyze some of the experiments in the cuprates which suggest that essential aspects of the physics are reflected in the intertwining of multiple orders---not just in the nature of each order by itself.  We also summarize and critique several theoretical ideas concerning the origin and implications of this complexity.  
\end{abstract}
\maketitle
\tableofcontents

\section{Introduction}
\label{intro}

%SAKUnconventional superconductors typically exhibit phase diagrams %SAK%with intrinsic complexity.  
Highly correlated electronic materials, and in particular those that exhibit unconventional superconductivity, have phase diagrams that
 are intrinsically complex.  Multiple distinct broken-symmetry phases occur as a function of parameters such as composition, pressure, 
%SAK
and magnetic field.  For example, there is a proximate antiferromagnetic state in the phase diagrams of superconductors such as cuprates, iron pnictides and chalcogenides, organics (both the quasi-1D TMTSF salts and the quasi-2D ET salts), and certain alkali-doped C$_{60}$ compounds. %SAK include prominent antiferromagnetic states. 
  It has become commonplace to describe %SAKstates
ordering tendencies such as superconductivity and antiferromagnetism as ``competing orders'' %SAK, as the static order parameters typically do not coexist.
since microscopic coexistence of the two broken symmetries  is relatively rare, and where they do coexist, one order manifestly suppresses the other.

The primary purpose of this colloquium is to emphasize a different perspective, taking the cuprates as a case study and focusing on the cooperative character of different orders.  We note that the temperature and energy scales associated with antiferromagnetism and superconductivity (SC) are comparable. Furthermore, while too much antiferromagnetism quenches superconductivity, experiments indicate that too little, in the form of residual (fluctuating) antiferromagnetism, is equally bad for superconductivity.  Increasingly, it has become clear that this is just the tip of the iceberg and that various other orders---charge-density wave (CDW), long period spin-density-wave (SDW), nematic, and possibly other forms of symmetry-breaking order---also occur with comparable onset-temperatures in a wide range of material parameters.  We present the case that the best way forward is to view these phenomena in terms of the ``intertwining'' of multiple orders.  %SAK, and we explain why we believe that alternative approaches, such as effective field theories, do not capture the key effects in a natural way.

A continuing conundrum in the cuprates concerns the nature of the normal state from which the superconductivity develops.  For a large range of carrier concentration, suppression of the superconductivity with temperature or magnetic field leads to the pseudogap %SAKphase
regime, a state with an ambiguous name that reflects an empirically well-defined set of electronic changes in the electronic structure whose underlying meaning is still much debated.%SAKcommunicates our continuing confusion about it.
\footnote{To see the range of ideas proposed to explain the pseudogap, one may turn to \textcite{norman-2005,lee-2006,rice-2012}.} There is evidence for various fluctuating or static order parameters within the pseudogap regime. %SAKphase.  We believe that the concept of intertwining orders can provide new insight; however, at present there is no controlled way (in two dimensions) to evaluate the fluctuating order parameters that should be relevant.  
A consistent theoretical description of such broad fluctuational regimes with multiple orders is possible in 1D and quasi-1D, from which some insight into the pseudo-gap can be gleaned, but no similarly compelling theory exists in 2D or 3D.
Nevertheless, %SAKthe essential features 
some of the essential features of the pseudo gap are addressed in our second theme, which is an exploration of a novel broken symmetry phase, the pair density wave (PDW), that intertwines CDW, SDW and SC orders.  There is increasingly compelling (although not yet  definitive) computational evidence that this novel phase exists robustly in the phase diagrams of simple models of strongly interacting electrons, and experimental evidence that it occurs in at least one cuprate SC, \LBCO.  More speculatively, we propose that the existence of such a ``parent'' phase which spontaneously breaks a large number of symmetries can be the  key to understanding broad aspects of the phase diagram in the sense that a large number of ``daughter''  phases can be viewed as partially melted versions of the parent phase, in which ``vestigial order'' still exists in the form of a smaller subset of  broken symmetries.

As the %SAKpair density wave
PDW is a new quantum phase of matter \cite{berg-2007,himeda-2002}, we need to define what we mean by it.  %SAKThe PDW
It is a state in which the superconducting order itself is spatially modulated in such a way that the uniform component is zero or nearly zero, but in which an oscillatory piece is strong.\footnote{We note that \textcite{chen-2004} used PDW to describe a different state, one of localized pairs.  In terms of broken symmetries, we classify the latter state as a CDW.}  This phase has unprecedented properties, of which the most readily experimentally identified are dynamical layer decoupling and anomalous sensitivity to disorder \cite{berg-2009b}. The PDW state can be viewed as a ``self-organized'' Larkin-Ovchinnikov (LO) state \cite{larkin-1964} but without the accompanying net magnetization.

%Because it has been so difficult to tease out clear evidence of what forms of broken symmetry are actually realized  from solutions of simple theoretical models of interacting electrons or from experiments on the cuprates, there has, until recently, been relatively little attention paid to the question of why so many different possible orders are present in delicate balance.   This is the focus of the present paper.  

The rest of the paper is organized as follows:
%In this paper, We begin this paper 
Section~\ref{elc} is a  qualitative discussion of the sorts of broken symmetry phases, especially somewhat less familiar electronic liquid crystalline phases, which can be expected in strongly correlated electron fluids.  In  particular, based on the analogy with the   liquid crystalline phases that occur in classical complex fluids, we offer some intuitive theoretical reasons to expect   intertwined orders to be an important generic feature of  broad classes of highly correlated electron systems.  From a somewhat different perspective, the ``landscape'' of possible ordered phases  that appear   at low temperatures in dynamical mean-field theoretic studies of strongly correlated systems \cite{kotliar-2005}, presumably reflects the same underlying physics.

In Sec.~\ref{explicit-models}, we discuss effective field theories of multiple interacting orders.    
%that following a brief survey of  we will outline the theoretical reasons to believe that intertwined orders are an important generic feature of a broad class of highly correlated electron systems, review some of the experimental  evidence that these considerations are relevant to the physics of the hole-doped cuprates (especially in the ``pseudo-gap regime'' of the phase diagram), and explore some of the important consequences of these observations.  Because it has been so difficult to tease out clear evidence of what forms of broken symmetry are actually realized  from solutions of simple theoretical models of interacting electrons or from experiments on the cuprates, there has, until recently, been relatively little attention paid to the question of why so many different possible orders are present in delicate balance.   This is the focus of the present paper.
%
%Much of the analysis we undertake is quite general. However, t
To give focus to the discussion, %and to explore  some particularly interesting consequences of the intertwining of various orders, 
we  %frame the discussion %starting from the %rviewpoint 
%supposition that %the there are
%in terms of  
consider the case in which there are two fundamental orders---a uniform 
($d$-wave) SC and a PDW.  From this starting point, other orders---notably CDW, % order and 
nematic, and charge 4e SC order---appear as composite orders.   
To some degree, the choice of which orders are treated as fundamental and which are derivative is a matter of convenience;  for instance, while it is %certainly straightforward
possible to describe CDW order as a composite, in regions of the phase diagram where no PDW condensation occurs it %would be 
is probably simpler to consider SC and CDW as the fundamental fields.  %However,  in terms of SC and CDW fields, PDW order would necessarily be described as an exotic fractionalized phase, so since PDW order 
 % if PDW order  plays a role in the phase diagram, it should be treated as fundamental.  
 In any case, SDW order, which is clearly an important part of the physics in  portions of the cuprate phase diagram, involves additional order parameter fields that we have not included to simplify the discussion.

Section~\ref{intertwined-microscopic}  reviews the results of a variety of theoretical studies of simple models of correlated electronic systems---mostly one version or another of the Hubbard model.  For the most part, we confine ourselves to a discussion of problems for which controlled analytical theory or arguably conclusive numerical solutions can be obtained.  In the first three subsections we study models that exhibit various general features of intertwined order.  In Sec.~\ref{models-pdw}, we focus on models which can be shown to have PDW ground states.  We also review recent, very illuminating variational results on the 2D $t$--$J$ model  which exhibit an astonishing near-degeneracy of a variety of different broken symmetry states---including a PDW phase---over a broad range of $t/J$ and doping concentration, $x$. In addition, we here briefly summarize a related approach to the problem \cite{lee-2014} which envisages a PDW state arising from ``amperian pairing'' of spinons in an underlying fractionalized phase.

Section~\ref{experiments} is a rather compressed summary of some of the most direct experimental evidence of the existence of a large variety of ordering tendencies in the cuprates.  The discussion here is more descriptive than analytic.  
%Much of this material has been reviewed more extensively elsewhere, including in a somewhat older  review by the present authors, \ref{rmp}.

In Sec.~\ref{Pseudogap}, we %SAKtouch on the mysterious and highly-debated pseudogap phase.  We 
summarize some of the spectroscopic features that are associated with the pseudogap, and in particular highlight the conflicting evidence, some of which is highly suggestive that the pseudogap is a fluctuational descendant of the $d$-wave superconducting gap, and some of which suggests  it arises from entirely distinct correlation effects, possibly associated with another form of order.  While we certainly do not resolve this debate, we do suggest that the existence of two distinct forms of superconducting order and/or order parameter fluctuations may provide a useful framework for resolving the apparent ``one-gap'' vs. ``two-gap'' dichotomy.

Section~\ref{tests} discusses several possible direct experimental tests which could unambiguously verify the existence of PDW order.  A more extensive discussion of most of the same points has appeared previously in \textcite{berg-2009b}.

In Sec.~\ref{discussion}, we consider some of the broader issues raised in the course of this colloquium.  The issue of whether it is reasonable to view the pseudogap  scale, $T^*$, as a crossover associated with the development of a local ``amplitude'' of the order parameter (or parameters) is discussed---but not resolved---in Sec.~\ref{preformedpairs}.  Using as illustrative examples the results from somewhat artificial model problems that are susceptible to controlled theoretical solution, a nontechnical physical discussion of known features of  the complex phase diagrams with intertwined orders is contained in Sec.~\ref{phase-diagrams}.  The role of dimensionality is explored, and special emphasis is placed on cases where partial melting of a highly ordered ground state can give rise to a variety of intermediate phases with vestigial order, and to cases (which probably, for technical reasons, are restricted to quasi-1D models) in which the ordered phases emerge on lowering $T$ from non-Fermi liquid ``normal'' states.
As discussed in Sec.~\ref{subsec-critique},
 %A closely related approach to the cuprate phase diagram, although from a more microscopic viewpoint, has  recently been presented by  \textcite{lee-2014} -- see, also, Sec.~\ref{amperian}. 
  various  versions of an  alternative approach, in which the existence of intertwined orders is associated with an assumed emergent higher (approximate) symmetry [{\it e.g.}, $SU(2)$, $SO(5)$ or $SO(6)$] that relates the various distinct forms of order, have been advocated by
 \textcite{zhang-1997}, \textcite{Efetov-2013}, and \textcite{hayward-2014}. While this approach has many attractive features, %we will (in Sec.~\ref{discussion}) 
 we point out what we consider to be rather general theoretical and phenomenological shortcomings of these scenarios.  We summarize our conclusions in Sec.~\ref{ineluctable}, %\textcolor{blue}{
 especially concerning the role of PDW order as a potential origin of intertwined orders in the cuprates.

\section{Electronic Liquid Crystal Phases}
\label{elc}

The intertwined orders observed in the cuprates appear to emerge with relatively light doping of a strongly correlated (Mott) insulating antiferromagnet.  A generic feature of short-range models of lightly doped Mott insulators, such as the the $t$-$J$ model, extended Hubbard models and others, 
is a strong  tendency toward phase separation in which the doped holes are expelled from  locally antiferromagnetic regions
 \cite{emery-1990,grilli-1991,emery-1993,vermeulen-1994,poilblanc-1995,misawa-2013}.

Motivated by these observations and by the discovery in 1995 by one of us of stripe phases in the Lanthanum family of cuprates \cite{tran95a}, two of us  introduced the concept of electronic liquid crystal phases, which we argued   are a general feature of the phase diagrams of strongly correlated systems \cite{kivelson-1998}.
\footnote{Electronic liquid crystal phases have also been studied in  the context of two-dimensional electron fluids in large magnetic fields \cite{fradkin-1999,Fradkin-2010,joynt-1996,balents-1996b}.}
The recent spectacular experimental discoveries of  diverse charge orders in the pseudogap phase of essentially all the cuprates (as well as in other materials, including evidence of electronic nematic order in  iron based \cite{fisher-2010} and heavy fermion superconductors \cite{matsuda-2011,Riggs-2014}) have generally validated the applicability of this concept.  (We will discuss these experiments in Sec.~\ref{experiments}.)

Also associated with local phase separation  is a tendency for the spins in the hole-poor regions to form local spin singlets, and  for the holes to pair as they aggregate \cite{white-1997}.  Hence, valence-bond crystals \cite{sachdev-2003} and uniform resonating valence bond (RVB) liquid states \cite{anderson-1987,kivelson-1987,lee-2006} are possible consequences of the same local physics. In particular, stripe (smectic) phases with a spin gap (which we discuss in section \ref{LE-ladders}) can be regarded as spatially non-uniform  RVB states.  At an intuitive level, the important point is that below a crossover temperature (which in the cuprates we would like to associate with the  crossover temperature to the pseudo-gap regime, $T^*$), the electron fluid should be thought of in terms of a fluid of spin singlets and small charged clusters, rather than of  electron-quasi-particles.  These clusters then behave in much the same way as the molecules in complex classical fluids, and consequently all sorts of ordering tendencies should appear in delicate balance below $T^*$, leading to  ``ineluctable complexity'' \cite{Fradkin-2012} of the phase diagram. 

Electronic liquid crystals are phases that spontaneously break translation and/or rotation symmetries; to make the analogy with classical liquid crystalline phases complete \cite{degennes-1993,chaikin-1995}, one might restrict attention to phases that remain conducting (fluid) despite the broken symmetries, although this condition is sometimes overlooked in common usage (as for a spin-nematic).   Examples range from multi-component CDW phases (which break translational symmetry in all directions), through stripe (or smectic) phases (which break translation symmetry along one direction and rotational symmetry), to  nematic phases (which  break spontaneously  only rotational invariance)  \cite{kivelson-1998}. 
For the electrons in a crystal, the symmetries available to be broken are the discrete translation and point group symmetries.  Unlike their classical cousins, electronic liquid crystalline phases can be strongly quantum mechanical.  From this comes the added richness of an interplay between the  order parameters associated with spatial symmetry breaking and intrinsically quantum orders, including superconductivity and magnetism.

Much has already been written concerning the microscopic mechanisms that lead to electronic liquid crystalline phases in strongly correlated electron systems;  for reviews, see \cite{kivelson-2003,Fradkin-2010,Fradkin-2012b,vojta-2009,fernandes-2014,Hu-2012}.  What constitutes  the principal focus of the following is an analysis of the way in which the existence of liquid crystalline phases leads to complex phase diagrams for electronic systems, just as it does   for classical complex fluids.  In particular, we will explore the important ways that superconductivity and liquid crystalline orders are naturally intertwined, so that the critical temperatures of various superconducting and charge ordered phases remain comparable to each other for a wide range of conditions.

\section{Field Theories of Intertwined Orders}
\label{explicit-models}

We now consider the effective field theories that describe a particularly interesting set of intertwined orders. In this context it is useful to consider both the relevant classical Landau-Ginzburg-Wilson (LGW) and  non-linear sigma models;  the former is an expansion in powers of the order parameter fields and so gives a reasonable description as long as the ordering is  weak, while the latter assumes a well developed local amplitude of the order parameter field, and focusses on the physics of the Goldstone modes (in the ordered phase) or the nearly-Goldstone modes in the fluctuational regime above $T_c$.  The number of order parameters that are potentially involved---both from theory and experiment in the cuprates---is dauntingly large, including as it does uniform SC, PDW, CDW, SDW, and nematic orders, at least.  

To keep the analysis manageable, we will ignore magnetism all together, and will treat both CDW and nematic orders as parasitic, deriving from a microscopic tendency to PDW order.  Starting with these we can  describe the uniform SC, the stripe ordered (CDW+SC), and the PDW states directly at mean-field level, while a pure CDW and a nematic phase, as well as various  more exotic phases including a charge $4e$ SC, can be obtained as states with ``vestigial order'' \cite{nie-2014} (involving composite order parameters) when fluctuation effects are treated carefully. For simplicity, we will also neglect the effects of quenched disorder (which is always a relevant perturbation as far as at least the CDW component of the ordering is concerned) and quantum fluctuations.  Of course, it is straightforward if tedious to include additional order parameter fields in the analysis, which is  necessary for explicit application to some materials.  

\subsection{Landau-Ginzburg Free Energy}
\label{LG-intertwined}

To begin with, we consider the LGW free energy density to quartic order in the fields where $\Delta_0$ is the uniform $d$-wave order parameter, ${\bm \Delta}_a = ( \Delta_{Q_a},\Delta_{-{Q_a}})$ is the complex spinor representing the two components of the PDW with ordering vector in the $a$ direction,
\begin{align}
F=&\frac {r_0} 2 |\Delta_0|^2 +\frac {r_Q} 2\Big[ |{\bm \Delta}_x|^2+|{\bm \Delta}_y|^2 \Big ] +\frac {\kappa_0} 2  |{\bm \nabla}\Delta_0|^2 \nonumber \\
+&\frac {\kappa_1} 2 \Big [ |\partial_x{\bm \Delta}_x|^2 + |\partial_y{\bm \Delta}_y|^2\Big ] +\frac {\kappa_2} 2 \Big [ |\partial_y{\bm \Delta}_x|^2 + |\partial_x{\bm \Delta}_y|^2\Big ] \nonumber \\
+& \frac u 4 \Big[ |\Delta_0|^2 +|{\bm \Delta}_x|^2+ |{\bm \Delta}_y|^2\Big ]^2 +\frac {\gamma_1} 2  |{\bm \Delta}_x|^2  |{\bm \Delta}_y|^2\nonumber \\
+&\frac {\gamma_2} 2 |\Delta_0|^2\Big[ |{\bm \Delta}_x|^2 + |{\bm \Delta}_y|^2\Big ] 
\nonumber \\
+&\frac{\gamma_3} 2 \Big[ |{\bm \Delta}_x^\dagger \tau_3{\bm \Delta}_x|^2+  |{\bm \Delta}_y^\dagger \tau_3{\bm \Delta}_y|^2\Big] \nonumber \\
+&\frac{\gamma_4}{2}\Big[({\bm \Delta}_x^\dagger \tau^+ {\bm \Delta}_y)({\bm \Delta}^\dagger_x \tau^- {\bm \Delta}_y)+{\rm c.c.}\Big] \nonumber\\
+& \frac {\gamma_5} 2 \Big[\big(\Delta_0^*\big)^2\big({\bm \Delta}_x^T \tau^-{\bm \Delta}_x +{\bm \Delta}_y^T \tau^- {\bm \Delta}_y\big)+ {\rm c.c.}\Big]
\label{LGW}
\end{align}
where $\tau_3$ and $\tau^\pm=(\tau_1\pm i \tau_2)/2$ are the three $2 \times 2$ Pauli matrices.
(This free energy was partly given by \textcite{agterberg-2008} using a different notation, and it is not even quite the most general possible form to this order.)  Because there are a total of 5 complex scalar fields involved, under fine tuned conditions $\kappa_j=\kappa$, $\gamma_j=0$, and $r_0=r_Q$ 
(8 conditions), this model has a large $O(10)$ symmetry. 

Away from such fine-tuned points, the symmetries of $F$ represent the microscopic symmetries of the system we have considered---gauge invariance, translational symmetry in the $x$ and $y$ directions, time-reversal symmetry, and various mirror and discrete rotational symmetries that exchange the $x$ and $y$ axes (and, at the same time, interchange ${\bm \Delta}_x$ and ${\bm \Delta}_y$).  

Even though we have greatly limited the number of ``primary'' order parameter fields treated explicitly, it is possible to study a variety of other order parameters as composites of the primary fields.  
 For instance, the unidirectional PDW ground-state with $\langle {\bm \Delta}_x\rangle \neq 0$, but $\langle \Delta_0\rangle = \langle {\bm \Delta}_y\rangle = 0$, breaks  gauge symmetry (it is a superconductor)   translational symmetry (it has a CDW component), and $C_4$ rotational symmetry  (it has a nematic component).  However, there are conditions in which, upon raising the temperature, such a PDW phase melts by a sequence of two transitions or more, at the lower of which gauge symmetry is restored but not translational symmetry, resulting in an intermediate unidirectional CDW phase which only melts at a second, higher transition temperature.  From this perspective, the CDW phase is viewed as a state with ``vestigial'' order, in the sense that it breaks some but not all of the symmetries that are broken by the fully ordered PDW ground state.  Naturally, if a set of primary fields have a non-zero expectation value, so do all products of those fields.  However, in a state with vestigial order, the expectation values of the primary fields vanishes, while certain composite order parameters still have non-zero expectation value \cite{berg-2009b,wang-2014b,lee-2014}.

While many forms of vestigial spin singlet orders can be envisaged in terms of the primary fields we have introduced, for the present discussion, 
three forms of order are most relevant \cite{berg-2009,berg-2009b}
\begin{itemize}
\item
{\em charge $4e$ uniform SC}, 
 with order parameters  
\begin{equation}
\Delta_{4e,a}=\Delta_{Q_a}  \Delta_{-Q_a}.
\label{4e}
\end{equation} 
which can exist in a nematic form ($|\Delta_{4e,x}|\neq |\Delta_{4e,y}|$) or in various rotational symmetry preserving ($s$, $d$, etc) forms.
\item
{\em 2Q-CDW} (unidirectional or bidirectional), whose order parameter is
\bea
\rho_{2{Q_a}} = \Delta_{-Q_a}^* \Delta_{Q_a}  
\label{CDW}
\eea
\item
{\em 1Q-CDW} (likewise, unidirectional or bidirectional) whose order parameter is
\bea
\rho_{Q_a} = \Delta_0^*\Delta_{Q_a} + \Delta_{-{Q_a}}^*\Delta_0 
\eea
\item
{\em Nematic order}, whose order parameter is
\bea 
{\cal N} = |\bm \Delta_x|^2-|\bm \Delta_y|^2
\eea
\end{itemize}
So, for example, a striped CDW phase modulated along the x direction arises if 
$\langle \Delta_{Q_a}\rangle=\langle \Delta_{0}\rangle=\langle \rho_{Q_y}\rangle = 0$ but $\langle \rho_{Q_x}\rangle\neq 0$,  
while a pure nematic phase arises if, in addition,  $\langle \rho_{Q_x}\rangle= 0$ but $\langle{\cal N}\rangle\neq 0$.  

\subsection{Non-linear sigma model---near multicriticality}
\label{all-together-now}

A non-linear sigma model description assumes the existence of a local amplitude of the order parameter which is well established, but which does not strongly distinguish between which form of superconducting correlations, uniform or modulated, is ultimately favored at long wave-lengths.   This amounts to assuming that one is in the vicinity of a multicritical point at which the critical temperatures of the the orders are equal. 

Were we to assume that the system were fine-tuned to the point of maximal symmetry, this would be an $O(10)$ non-linear sigma model, with many interesting features, including (by selectively breaking subsets of this symmetry) all the interesting features of $SO(5)$, $SO(6)$ and $SU(2)$.  Instead, we will assume only the actual symmetries of the problem, but treat the problem under the assumption that the  temperature dependence of the amplitudes of the various order parameters can be neglected over a suitable range of temperatures;  this is an innocuous assumption for the purposes of studying critical phenomena, but  is more problematic if we are interested (as we are) in the properties of the system over an extended range of $T$.  It is analogous to treating spin antiferromagnetism in an approximation that assumes the existence of a fixed magnitude local moment \cite{chakravarty-1988,chakravarty-1989} or the phase fluctuations of a superconductor with fixed magnitude local pairing strength \cite{carlson-1999,eckl-2002}.  This is a highly non-trivial assumption---it implies that the ``mean-field transition temperature'' (itself an ill-defined concept) for both the uniform SC and PDW order is higher than the temperatures at which any of the ordering phenomena of interest occur, and that therefore (again in a somewhat ill-defined sense) the local magnitude of the corresponding order parameters are well defined over a broad range of temperatures which extends well above any observed $T_c$.  We will discuss the issue of whether this assumption is consistent with experiment in the cuprates in Sec.~\ref{preformedpairs}.
 
Leaving aside worries about the range of validity of such an approximation, there is still an issue concerning the role of various possible patterns of discrete symmetry breaking that can arise.  To simplify the discussion,  {\it i.e.} to avoid the many special considerations needed to be completely general, we will explicitly assume that time-reversal and inversion symmetry remain unbroken, which implies that the amplitudes at $\pm Q_a$ are the same,  $|\Delta_{+Q_a}|^2 =|\Delta_{-Q_a}|^2$.  We can then define five phase fields as $\Delta_0 \equiv |\Delta_0|e^{i\theta_0}$ and $\Delta_{\pm Q_a} \equiv |\Delta_{Q_a}|e^{i(\theta_a \pm \phi_a)}$, and a real scalar field (representing possible Ising-nematic order), ${\cal N} = |\Delta_{Q_x}|^2 - |\Delta_{Q_y}|^2$.  In any isotropic phase (where ${\cal N}=0$), the effective Hamiltonian in terms of the phase fields takes the form
\begin{align}
H=&\frac {K_0} 2 {(\bm \nabla}\theta_0)^2 + \frac {K_{1}} 2
  \Big[
(\partial_x\theta_x)^2+ 
(\partial_y\theta_y)^2\Big]
 \nonumber\\ 
+&\frac {K_2} 2 \Big[( \partial_y\theta_x)^2+ ( \partial_x\theta_y)^2\Big] +\frac {K_3} 2 \Big[(\partial_x\phi_x)^2+ (\partial_y\phi_y)^2\Big] \nonumber\\
+&\frac {K_4} 2 \Big[(\partial_y\phi_x)^2+ (\partial_x\phi_y)^2\Big]-\tilde V\cos(2\theta_x-2\theta_y) \nonumber \\
-& V\Big[\cos(2\theta_0-2\theta_x)+\cos(2\theta_0-2\theta_y)\Big]
\label{NLSM-U1s}
\end{align}
From Eq.~\eqref{CDW} it is apparent that $2\phi_a$ is the phase of the $2Q$ CDW component of the order.  It is important to remember that while the CDW and SC phases seem to be totally uncoupled, they are related through the condition that $e^{i(\theta_x\pm\phi_x)}$ and $e^{i(\theta_y\pm\phi_y)}$ 
are single valued.  As we will see in Sec.~\ref{phase-diagrams}, this controls the nature of the topological excitations (vortices).  On the other hand, the superconducting phases, $\theta_0$, $\theta_x$ and $\theta_y$, are coupled to each other by higher order Josephson-like terms---the cosine terms in Eq.~\eqref{NLSM-U1s}.  This leaves us with three $U(1)$'s and two $\mathbb{Z}_2$ symmetries, the latter the remnant of the superconducting relative phases.

In an Ising nematic phase, the effective phase stiffnesses  are anisotropic, $K_j \to K_{ja}({\cal N})$, where $[K_{jx}({\cal N})-K_{jy}({\cal N})]=-[K_{jx}(-{\cal N})-K_{jy}(-{\cal N})]$, similarly $V \to V_{ja}({\cal N})$ with $[V_{x}({\cal N})-V_{y}({\cal N})]=-[V_{x}(-{\cal N})-V_{y}(-{\cal N})]$, and where $H$ must be augmented by an effective Hamiltonian for ${\cal N}$.  Deep in a nematic phase, with ${\cal N}>0$, we can simply assume that the fluctuations of $\phi_y$ and $\theta_y$ are so violent that these fields can be ignored, leaving an effective model (which we will study in Sec.~\ref{phase-diagrams}) with $U(1)\times U(1) \times \mathbb{Z}_2$ symmetry.

\section{Intertwined order in  Hubbard models }
\label{intertwined-microscopic}

\subsection{Lightly doped antiferromagnetic insulators}
\label{smallx}

The Hubbard model and the related $t$-$J$ models are widely thought to  capture the essential physics of a class of highly correlated systems.  On a two-dimensional square lattice, when the band parameters ({\it e.g.} the ratio between the first and second neighbor hopping matrix elements $t'/t$) are adjusted to reproduce the salient features of the experimentally measured Fermi surface, these models may be sufficiently ``realistic'' that their properties can be compared with experiment in the cuprates \cite{scalapino-2012,emery-1987,varma-1989}.

In the weak coupling limit, $U/t \ll 1$, an asymptotically exact solution of the Hubbard model \cite{Raghu-2010} is possible: down to exponentially low temperatures of order $T_{\rm SC} \sim 4t \exp[- \alpha (t/U)^2]$, the model exhibits Fermi liquid behavior, $\alpha$ is a number of order 1 which depends on details of the band-structure and the value of electron concentration per site, $n$.  Below this temperature, the system exhibits $d_{x^2-y^2}$ superconductivity  for a broad range of  $n$ in the neighborhood of $n=1$ with   $T_c \sim T_{\rm SC}$.  Although the nature of the superconducting state itself is reminiscent of the superconducting state in the cuprates, this is where the resemblance ends.  In particular, there are no strong fluctuation effects to give rise to pseudo-gap phenomena, and no trace of competing orders of any sort.  This is generic behavior for any weakly interacting Fermi fluid in more than one dimension.

In the strong coupling limit, $U/t \gg 1$ or $t/J \gg 1$, recent numerical studies have confirmed \cite{Li-2012} what was long believed \cite{nagaoka-1966}, that both models exhibit fully polarized ferromagnetic metallic phases for a 
broad range of $n$ near $n=1$.  Again, there are no competing orders.

Thus, to the extent that intertwined, or even conventional competing orders are features of the theoretically expected landscape, they must arise exclusively at intermediate coupling.  For intermediate coupling, where $U$ is of order the bandwidth, $U\sim 8t$ or $J/t \sim 1/2$, there have been many approximate (mean-field) and numerically implemented  approaches to the problem.  Various different conjectured phases have been found in different studies---already evidence that no single pattern of broken symmetry is strongly favored.

For $n=1$ ($x=0$), $U$ of order the bandwidth and $t'$ not too large, the Hubbard model has an antiferromagnetically-ordered, insulating ground-state \cite{lin-1987,chakravarty-1989,arovas-1988}.  In the decade following the discovery of superconductivity in the cuprates, the issue of how the system evolves with weak doping, $|x| \ll 1$, was one of the most studied problems in condensed matter physics.  The problem is complex since there is  inherent frustration between the tendency 
 to maintain local antiferromagnetic correlations  and the doped hole itineracy  Three possible scenarios have been considered.

One frequently occurring possibility \cite{emery-1990,emery-1993} is that this evolution is discontinuous, leading to macroscopic phase separation into regions of undoped antiferromagnet  where the antiferromagnetic exchange is unfrustrated, and critically doped regions with $x=x_c(U/t)$, where the zero point energy of the doped holes is  dominant.  Phase separation has been shown to occur in the limit of large spatial dimension  $d\gg 1$ \cite{carlson-1998}, large spin $S \gg 1$ \cite{auerbach-1991b}, in certain large $N$ generalizations of the problem \cite{auerbach-1991a}, and  for $U/t \gg 1$ \cite{emery-1990,Li-2012,hellberg-1997,misawa-2013}.  
Given that the ground state at $x=0$ is a N{\'e}el state, if one assumes, consistent with an RVB scenario, that the lightly doped system corresponds to a doped  spin liquid \cite{anderson-1987,kivelson-1987,rokhsar-1988,fradkin-1988,anderson-2004,lee-2006}, the transition between these two states must be first order, and thus there must be a two-phase region for small enough $x$.

Another possibility is some form of  local phase separation---especially stripe formation---which is driven by more or less the same local energetic considerations.  The earliest such proposals \cite{zaanen-1989,schulz-1990,machida-1989} suggested insulating stripes, {\it i.e.}\ unidirectional charge density waves with a doping dependent period such that $x_s$, the density of doped holes per site perpendicular to the charge ordering vector, is fixed at $x_s=1$.  Later proposals \cite{white-1998a,troyer-1995,nayak-1997} based on various  approaches to the 2D $t$-$J$ and Hubbard models suggested conducting stripes, with $0 < x_s < 1$.  Some of these studies found evidence that the doped holes in the conducting stripes are strongly paired, and possibly superconducting.   Since in the presence of long-range Coulomb interactions (not included in Hubbard-like models), macroscopic phase separation of charged particles is not possible, where the short-range interactions tend to produce phase separation, Coulomb-frustrated phase separation generically results in CDW order of a sort that is difficult to distinguish from direct forms of local phase separation, and similarly, where the short-range interactions favor stripe formation, long-range Coulomb forces will tend to shift the period, with the associated tendency to turn insulating  to conducting stripe phases.  Thus, in practice, the difference between cases 1 and 2 is not physically significant.

The third possibility is one form or another of uniform phase with coexisting antiferromagnetic order \cite{zhang-1997,demler-2004}.   Note that a discontinuous drop of the sublattice magnetization would necessarily imply a first order transition and hence phase separation.  (For references see \textcite{misawa-2013}).

\subsection{Intertwined orders in model quasi 1D systems}
\label{LE-ladders}

Powerful field theory methods (bosonization) permit an essentially complete understanding of the  long-distance properties of interacting electrons in 1D, while efficient numerical methods (especially DMRG) permit the short-distance microscopic physics of specific models, even of rather complicated multi-leg ladders, to be treated reliably.  (For a review of bosonization and Luttinger and Luther-Emery liquids see chapter 6 of \cite{fradkin-2013}). By matching results at intermediate length scales it is possible to obtain an essentially complete theoretical understanding of strongly-correlated systems in 1D.  Moreover, so long as interchain couplings are sufficiently weak, various forms of inter-chain mean-field theory allow these results to be extended to quasi-1D systems.  Taken literally, these models are relevant only to the properties of quasi 1D materials, but in many cases, results in this ``solvable'' limit give qualitative insight into the behavior of fully 2D or 3D highly correlated electron systems
 \cite{kivelson-1998,carlson-2000,emery-2000,vishwanath-2001,granath-2001,mukhopadhyay-2001,carr-2002,essler-2002,jaefari-2010,teo-2014}.  In particular, this is the only well understood class of problems in which various ordered phases emerge from non-Fermi liquids.

It is known from density-matrix renormalization group (DMRG) studies \cite{noack-1997} and  bosonization theories \cite{balents-1996,wu-2003,controzzi-2005} that $t$-$J$ and Hubbard-type models on a  two-leg ladder  have a broad regime in which they have a spin gap $\Delta_s\approx J/2$ and $d$-wave-like superconducting correlations.  DMRG studies find \cite{noack-1997,troyer-2001,white-2002} that the spin gap decreases monotonically with doping $x$,  and vanishes at a critical doping  $x_c \approx 0.3$.  The upshot is that the two-leg ladder is in a Luther-Emery (LE) liquid phase whose effective field theory consists of a field $\phi$, describing the phase of the incommensurate CDW amplitude $\rho_Q(x) \sim \exp[i\sqrt{2\pi} \phi(x)]$ (with  $Q=2k_F$) and its dual field $\theta$,  describing the phase of the superconducting order parameter $\Delta_0(x) \sim \exp[i\sqrt{2\pi} \theta(x)]$.  The  field theory contains two important parameters---the  charge Luttinger parameter, $K_c$, and  the charge  velocity, $v_c$.   Both are (complicated) functions of the microscopic parameters of the ladder.   DMRG studies showed that $K_c\to 2$ as $x \to 0$ and decreases with increasing doping,  reaching the value $K_c = 1/2$  at $x_c \approx 0.3$ (where $\Delta_s \to 0$).  For $T \ll \Delta_s$, the SC and CDW susceptibilities obey the scaling laws $\chi_{\rm SC}\sim \Delta_s /T^{2-K_c^{-1}}$ and $\chi_{\rm CDW} \sim \Delta_s/T^{2-K_c}$. 
 
Thus, an interesting quasi 1D model to analyze is an array of spacing $a$ of  two-leg Hubbard ladders  \cite{emery-2000,arrigoni-2004}.  At energies low compared to the single ladder spin-gap, the array is represented by  a set of CDW phase fields $\{ \phi_j \}$ and their conjugate SC phase fields $\{ \theta_j \}$.  Although microscopically there are many  local couplings between neighboring two-leg ladders which are allowed, if the couplings are weak compared to $\Delta_s$, most of these are irrelevant.  In particular, electron tunneling, that normally drives the quasi-1D system into a 2D Fermi liquid  state, is suppressed.   The marginal or potentially relevant inter-ladder interactions are: a)  $\mathcal{J}_{\rm SC}$, the Josephson coupling between the SC order parameters of nearest neighbor ladders ({\it i.e.} Cooper pair tunneling) which locks $\theta_j(x)$ on neighboring ladders, b) $\mathcal{J}_{\rm CDW}$, the coupling of the CDW order parameters of neighboring ladders which  locks $\phi_j(x)$ on neighboring ladders, and c) $\tilde V$, inter-ladder forward scattering interactions (which are strictly marginal operators).  

Models of this type have been studied extensively using a variety of techniques \cite{carlson-2000,arrigoni-2004,emery-2000,vishwanath-2001,jaefari-2010}.  In the range in which there is a spin gap, SC and CDW orders compete.  The resulting state is (roughly) determined by which coupling is more relevant.  The scaling dimensions of the Josephson coupling and of the CDW coupling are $D_{\rm SC}=1/K_c$ and  $D_{\rm CDW}=K_c$, respectively. 
For small $x$, $D_{\rm SC} \approx 1/2$ and $D_{\rm CDW}\approx 2$, so  the Josephson coupling is strongly  relevant while the CDW coupling is barely relevant.   Conversely, for larger values of $x$ the CDW coupling is more relevant than the Josephson coupling. At an intermediate value of $x$ (corresponding to $x_P \simeq 0.1$)  $K_c=1$ at which point $D_{\rm CDW}=D_{\rm SC}=1$ so the two orders are equally relevant.  This balance can be affected, as well, by the marginal interactions, $\tilde V$ \cite{emery-2000,vishwanath-2001}. 

\begin{figure}[t]
 \begin{center}
 \includegraphics[width=0.4\textwidth]{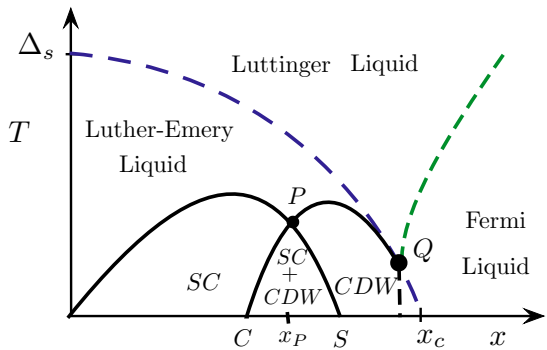}
 \end{center}
 \caption{(color online)  Schematic phase diagram (temperature $T$ vs doping $x$) for an array of weakly coupled two-leg Hubbard ladders.  The long-dashed curve is the spin gap, which vanishes at $x_c\approx 0.3$, and which also indicates a crossover temperature  to a LE liquid regime.  The  short-dashed line indicates a crossover scale to a higher dimensional Fermi liquid (FL) regime.  $x_P\approx 0.1$ is the point at which $K_c=1$ and $P$ is the corresponding tetra-critical point discussed in the text.   $C$ and $S$ are quantum critical points, between which SC and CDW orders coexist. 
We have shown $Q$ as a trictitical point, below which the transition   (dark short dashed line) between the CDW and the FL becomes first order, although there are other possibilities here. For details, see text.}
 \label{fig:phasediagram1}
 \end{figure}

The generic phase diagram \cite{arrigoni-2004,jaefari-2010}  for weakly coupled 2-leg Hubbard ladders is shown in Fig.~\ref{fig:phasediagram1}.  (See also Fig.~5.)  One thing  to note about this phase diagram is that the ordered phases emerge from a non-Fermi-liquid ``normal'' state  which for $T>\Delta_s$ consists of effectively decoupled  LE liquids.   For $K_c$ not too close to $K_c=1$, either SC or CDW order is dominant.  The corresponding critical temperatures  can be estimated from dimensional crossover mean field theory, and are found to have a power-law dependence on the inter-ladder couplings with an overall scale set by the spin gap \cite{carlson-2000,arrigoni-2004}
 \begin{equation}
 T_{\rm SC} \sim \Delta_s \left({\mathcal{J}_{\rm SC}}\right)^{\alpha_{\rm SC}}, \; 
 T_{\rm CDW} \sim \Delta_s \left({\mathcal{J}_{\rm CDW}}\right)^{\alpha_{\rm CDW}},
 \label{eq:Tcbounds}
 \end{equation}
where the exponents are $\alpha_{\rm SC}= K_c/(2K_c-1)$ and $\alpha_{\rm CDW}=1/(2-K_c)$, respectively.  $T_{\rm CDW}$ vanishes as $x\to x_c$ due to the vanishing of $\Delta_s$;  for larger $x$, the most relevant 
 interladder coupling is single-particle hoping which leads to a crossover to a higher-dimensional Fermi liquid regime that is stable to exponentially low temperatures.  The vanishing of $T_{\rm SC}$ in proportion to $x$ as $x\to 0$ involves additional considerations associated with the approach to the correlated (Mott) insulating state at $x=0$, as discussed in \onlinecite{arrigoni-2004}.
 
For $K_c\approx 1$, there is a multi-critical point (shown as $P$ in Fig. 1) at which $T_{\rm SC}$ and $T_{\rm CDW}$ meet.  The dimensional crossover mean-field theory predicts that the 2D array has a SC phase and a CDW phase, but it does not tell us if these phases are separated by a first order transition \cite{emery-2000,carr-2002} or if there is a phase in which SC and CDW orders coexist. 
It is a special (and likely non-generic) feature of the two-leg ladder that the CDW coupling becomes marginal for the same value $x_c$ where the spin gap vanishes. In the phase diagram of Fig. 1 we  allowed for a tricritical point, $Q$, and a first order transition out of the CDW state into a FL state (which has no spin gap), to remove this accidental feature.

To address this problem, \textcite{jaefari-2010}  derived an effective field theory, a nonlinear sigma model in $2+1$ dimensions that describes the fluctuations of both the SC and the CDW order parameters of the ladder array under the special fine-tuned conditions that $K_c=1$ and $\mathcal{J}_{\rm SC}=\mathcal{J}_{\rm CDW}$.  Although the microscopic model has only a global $U(1) \times U(1) \simeq O(2) \times O(2)$ symmetry, they showed that under these special circumstances there exists an enlarged $O(4)$ symmetry which unifies the CDW and SC orders.  Thus, for $K_c$ close to 1 and $|\mathcal{J}_{\rm SC}-\mathcal{J}_{\rm CDW}|\ll W$, where $W$ is a high-energy cutoff for the charge sector of the ladder \cite{carlson-2000},  the critical fluctuations near this fine-tuned multicritical point ($P$ in Fig. 1) can be described by the effective action 
\footnote{Here we have omitted the spatial anisotropy in the gradient terms of the 2D system since under the RG the anisotropy is a redundant operator \cite{affleck-1996}.}
for an $O(4)$ nonlinear sigma model (NLSM) \cite{jaefari-2010} 
 \begin{equation}
 S[n]=\frac{1}{2g_0 } \left(\partial_\mu n\right)^2+w \ \partial_\mu n_a O_{ab} \partial^\mu n_b+ h\ n_a O_{ab} n_b
 \label{O4NLSM}
 \end{equation}
where $n_a$ is a four-component order parameter subject to the constraint $n_an_a=1$, where $n_1/n_2 =\tan(\theta_{\rm SC})$, $n_3/n_4=\tan(\phi_{\rm CDW})$, and $\mu=0,1,2$ are the  the space-time indices.  
  The first term of the action of Eq.~\eqref{O4NLSM} has an effective coupling constant $g_0\simeq \left[8\pi(\Delta_s/v_c)  (W/(\mathcal{J}_{\rm SC}+\mathcal{J}_{\rm CDW})a) \right]^{1/2}$.  The last two terms  represent the breaking of $O(4) \to O(2) \times O(2)$. Here $O_{ab}={\rm diag} (1,1,-1,-1)$ is a diagonal $4 \times 4$ matrix, $w \propto (K_c-1)$ and $h \propto \Delta_s (\mathcal{J}_{\rm SC}-\mathcal{J}_{\rm CDW})/2W$.\footnote{\textcite{jaefari-2010} also found  a topological term that could  lead to a deconfined quantum critical (DQC) point, but which in this case is inaccessible  due to the presence of spatially anisotropic perturbations \cite{senthil-2006}.}

The nature of the phase diagram, and in particular the existence of the tetracritical point, $P$, was obtained by \textcite{jaefari-2010} from a standard analysis of the (quantum) NLSM in 2D  (and its relation with the LGW $\phi^4$ field theory formulation), both at finite temperatures and at zero temperature.  Of course, in 2D at finite temperature the designations SC and CDW refer to phases with quasi-long-range order, while they are long-range ordered in 3D.  Both of the symmetry breaking terms, $w$ and $h$, are perturbatively relevant, so the tetracritical point generically has only the $U(1)\times U(1)$ symmetry of the microscopic model.  Consequently, even in 2D, $P$ occurs at finite temperature. However, under the doubly fine-tuned circumstances $w=h=0$, the resulting $O(4)$ symmetry implies that in 2D the tetracritical point is suppressed to $T=0$.  The emergent $O(4)$ symmetry at $w=h=0$ is similar to such symmetries that arise in the  $SO(5)$ theory of antiferromagnetism and superconductivity of \textcite{zhang-1997} and in the $O(6)$ of \textcite{hayward-2014} and $SU(2)$ of \textcite{Efetov-2013} of  CDW  and SC order, which we will discuss in Sec.~\ref{subsec-critique}.
 
In summary, this analysis shows that the higher symmetries are, in general, {\em not emergent} symmetries but instead are fragile. In spite of that, systems of this type in general have intertwined orders with complex phase diagrams with  several quantum critical points under their dome(s) \cite{kivelson-1998}.   On the other hand, this analysis shows that (in this case, at least) these quantum critical points cannot be regarded as the origin of the SC (or the CDW) phase, 
contrary to what is often assumed in the literature of {\hty}.

\subsection{Variational results for the 2D $t$-$J$ model}
\label{variational-t-J}

In a recent  paper, \textcite{corboz-2014} have found the ``best'',  to date, variational solution of the 2D $t$-$J$ model at intermediate coupling on the square lattice, in the technical sense that they have obtained the lowest variational energy.  While there is always the danger with variational studies that the true ground state could have properties that are incompatible with the assumed form of the states considered, the only obvious prejudice of the present study is that it favors states with relatively lower ``quantum entanglement.''  Given this and the large number of variational parameters involved, it is quite plausible that the present results can be taken at face value.

Corboz {\it et al.}\ find that for the range of parameters studied ({\it i.e.}, $J/t$ between 0.2 and 0.8 and doped hole concentration  $x\equiv 1-n$ between 0 and 0.16) the states listed below all have energies that are, to a high degree of accuracy, equal to each other: 
1) A uniform $d$-wave superconducting phase (SC) corresponding to $\langle \Delta_0\rangle \neq 0$, and $\langle \Delta_{Q_a}\rangle=0$.  For $x < x_c$ (where $x_c = 0.1$ for $J/t=0.4$) this has coexisting antiferromagnetic N\'eel type magnetic order.   
2) A state with coexisting $2Q_x$ charge density wave and uniform $d$-wave superconducting order (CDW+SC).  This is a Òstriped-superconducting-phaseÓ with  $\langle \Delta_0 \rangle \neq 0$,             $\langle \rho_{2Q_x} \rangle \neq 0$, and $\langle \rho_{2Q_y} \rangle= 0$ which spontaneously breaks translational and $C_4$ lattice rotational symmetry.  
3) A unidirectional PDW phase with no uniform SC component,  i.e. with $\langle \Delta_0 \rangle = 0$, $|\langle \Delta_{Q_x}\rangle|=|\langle\Delta_{-Q_x}\rangle |\neq 0$ (from which it follows that  $\langle \rho_{2Q_x}\rangle \neq 0$), and $|\langle \Delta_{\pm Q_y}\rangle |= 0$.
Both the CDW+SC and the PDW states are found to break spin-rotational symmetry, as well, through formation of unidirectional SDW order with a modulated amplitude such that the superconducting component of the order has its maximum amplitude at nodes of the SDW order and is minimal (vanishes in the case of the PDW) where the SDW amplitude is maximal.  (To describe the magnetic components of these orders, we would need to include additional fields---which we have neglected for simplicity elsewhere in this paper.)
	
Naturally, it is not true that these three distinct phases are exactly degenerate; the CDW+SC phase achieves the lowest variational energy.   However, the ground-state energy per site of the CDW+SC is lower than the PDW only by roughly $\Delta E = 0.001\,tx$, and than the uniform SC by roughly $\Delta E = 0.01\,tx$. These differences are so small that it is not clear that they are significant (within the accuracy of the variational ansatz), and in any case one would expect that small changes to the model could easily tip the balance one way or the other.  At the rough intuitive level, this near degeneracy reflects the fact that locally, all three phases look pretty similar in that they all look like a uniform $d$-wave superconductor, with or without coexisting antiferromagnetism depending on the local doped hole concentration.  

A few other aspects of the results are significant as well:  1) The periodicity of the CDW order for either of the striped phases is not determined by Fermi surface nesting features; rather the preferred density of holes per unit length of stripe, $n_s$, is a function of the value of $J/t$ ({\it i.e.}, is determined by the strength of the interactions), ranging from about $n_s=0.35$ for $J/t=0.2$ to $n_s=1$ (corresponding to insulating stripes) for $J/t=0.8$.  
2) The SDW component of the order suffers a $\pi$ phase shift across the row of sites at which the CDW order is maximal;  thus, for even period CDW order, the SDW period is twice that of the CDW (which is the same as the period of the SC order in the case of the PDW), while for odd period CDW order, the SDW period is equal to that of the CDW.     
3) In the context of the cuprates, there has been considerable discussion of whether striped states or checkerboard states (bidirectional CDW states that preserve the $C_4$ rotational symmetry of the underlying lattice) are preferred; an earlier version of this  calculation \cite{corboz-2011}
indicated that the checkerboard phase is never preferred.  
An insulating diagonal striped phase (with $n_s=1$) was also found to have relatively low variational energy, but never competitive with the vertical stripe phases. 

\subsection{%Models with PDW ground states
%SAK
Pair density waves (PDW) in model systems}
\label{models-pdw}

There is no material system in which we know {\em for certain} that PDW order occurs.  Thus, it is worth addressing as a point of principle whether this phase occurs in any theoretically tractable microscopic model.

\subsubsection{PDW phases in Hubbard-like ladders}
\label{pdw-quasi1D-ladders}

We already noted in section \ref{LE-ladders} that DMRG studies of Hubbard and $t$-$J$ two-leg ladders have revealed that these systems have a spin gap over a significant hole doping range, and that they exhibit strong superconducting correlations.  What is less widely recognized is that, in many cases, what is formed is a 1D version of a PDW, in which uniform SC order parameter correlations fall exponentially with distance, while there exist charge $2e$  finite momentum ($2k_F$) superconducting correlations 
($\Delta_{2k_F}$) and uniform charge $4e$ ($\Delta_{4e}$) correlations which exhibit power-law fall-off of spatial correlations and a divergent $T=0$ susceptibility.  
 
To begin with, we consider a highly asymmetric ladder -- the Kondo-Heisenberg (KH) chain.
  Its 3D cousin is often used as a model of heavy-fermion systems.  The 1D version consists of an interacting electron gas [a Luttinger liquid (LL) gapless both in its spin and charge sectors] and a spin chain (with exchange coupling $J_H$), coupled to each other by the Kondo exchange interaction $J_K$.  The KH chain has been studied by many authors using diverse methods \cite{zachar-1996,sikkema-1997,coleman-1999,zachar-2001,zachar-2001b,Berg-2010}.  From DMRG studies \cite{sikkema-1997} it is known that there is a broad range of parameters $J_H/J_K$ and electron densities in which the KH chain has a spin gap, corresponding to the ``Kondo-singlet'' regime of the heavy-fermion literature. 

However, both DMRG stimulations \cite{Berg-2010} and bosonized effective theories \cite{zachar-2001,zachar-2001b,Berg-2010} show that  in the spin-gap regime,  the correlators of {\em all} fermion bilinears are short ranged, including the N{\'e}el order parameter of the spin chain, the SDW order parameter of the LL, all the fermionic pair fields that describe possible uniform SC order parameters of the LL (both singlet and triplet), as well as the particle-hole CDW order parameters of the LL. 
Specifically,  \textcite{Berg-2010} showed (using both DMRG simulations and bosonization) that  the most prominent long-range SC correlations involve 
$\Delta_{4e}$ 
and 
 a composite PDW order parameter of the form 
$ \Delta_{2k_F}= {\bm \Delta} \cdot {\bm N}$, where ${\bm \Delta}$ is the uniform {\em triplet} SC order parameter of the LL and ${\bm N}$ is the N{\'e}el order parameter of the spin chain.  The PDW order parameter inherits the ordering wave vector $Q=\pi/a$ (where $a$ is the lattice spacing of the spin chain)  of the short-range N{\'e}el correlations of the spin chain. Similarly, in this phase there are four-fermion CDW order parameters with power-law correlators.  Thus, the Kondo singlet regime of the KH chain is a PDW, 
that cannot be described by a conventional condensation of Cooper pairs with finite momentum. 

 Turning to the more usual (symmetric) two-leg ladder \textcite{jaefari-2012},  extending the results and methods of \textcite{wu-2003}, found that, in addition to $d$-wave superconductivity coexisting with stripe charge order, 
these ladders also have commensurate PDW phases. In the weak coupling regime, in which bosonization is most accurate, the PDW phases arise as follows: The electronic structure of a non-interacting two-leg ladder has a bonding and an anti-bonding band. For certain values of the electron density (and for strong enough repulsive interactions) the bonding band is at a commensurate filling and umklapp processes open a charge gap on that  band. 
Except for the special case of a half-filled bonding band,  for general commensurate filling this Mott insulator is a commensurate charge-density-wave (with ordering wave vector $Q/2$) that coexists with a gapless spin sector whose low-energy behavior is that of a spin-1/2 antiferromagnetic chain with N{\'e}el quasi-long range order characterized by  wave vector $Q$. In this regime, the spin sectors of the bonding and anti-bonding bands are coupled by a Kondo-type exchange coupling, which  is a marginally relevant perturbation that drives the ladder into a state with a full spin gap.

Thus, in this regime the low-energy degrees of freedom of the two-leg ladder are quite similar to those found in the KH chain.  Indeed, \textcite{jaefari-2012}  found two distinct SC phases: a) a uniform $d$-wave SC which coexists with a CDW wave and b) a commensurate PDW phase whose order parameters are composite operators of the uniform triplet superconductor of the anti bonding band and the SDW (``N{\'e}el) order parameter for the bonding band.  Here, too, both bilinear order parameters are short ranged but their product has power law correlations.

\subsubsection{Occurrence of the PDW in mean-field theory}
\label{pdw-mft}

Inhomogeneous superconducting states closely related to the PDW have been found in several mean-field theories of superconducting states in strongly correlated systems  \cite{himeda-2002,raczkowski-2007,yang-2008b,Loder-2010,zelli-2011,lee-2014,soto_garrido-2014}.  \textcite{Loder-2010,Loder-2011}  recently showed that it is possible to obtain a PDW SC state (with and without an SDW component) using a BCS mean-field theory. 
Their work  
uses a $t-t'$ tight binding model with a ``realistic'' Fermi surface appropriate for the cuprate superconductors and an {\em attractive} nearest neighbor interaction $V$.
Naturally, since no generic band-structure has a divergent PDW susceptibility, Loder and coworkers find that this state only occurs for  fairly large attractive interactions.  On the other hand, for such large values of the attractive interaction, the applicability of BCS mean-field theory is questionable. This problem is well known in the conventional Larkin-Ovchinnikov state \cite{larkin-1964} and the related Fulde-Ferrell 
  time-reversal-breaking SC state \cite{fulde-1964}.  [Similar issues arise in spin-imbalanced Fermi gases 
\cite{radzihovsky-2008,radzihovsky-2011}.] 

Putting these caveats aside,  the results of \textcite{Loder-2010,Loder-2011} are quite suggestive.  As expected,  
for small $V$ they find uniform $d$-wave superconductivity. However, in the vicinity of $x=1/8$ hole doping, for a range of interactions $1\lesssim V \lesssim 3$ they find a PDW phase which coexists with charge and spin stripe order, while for $V\gtrsim 3$, an intertwined charge and spin stripe state survives (without superconductivity).  Similar results were found earlier in variational Monte Carlo simulations of the $t$-$J$ model by \textcite{himeda-2002} and by \textcite{raczkowski-2007},  and in a mean-field  theory of the $t$-$J$ model by  \textcite{yang-2008b}. 

A potentially significant feature of these mean-field results is the nature of the quasiparticle spectrum in the PDW state.  In a PDW state without spin-stripe order, there is a ``pseudo-Fermi surface'' of Bogoliubov quasiparticles with multiple pockets of increasing complexity as the ordering period  changes \cite{baruch-2008,berg-2008a,berg-2009b,Loder-2011,zelli-2011,lee-2014}. However, the presence of even a small amount of uniform $d$-wave SC component gaps out all these pockets, leaving only the usual nodal states of a $d$-wave superconductor.  Coexisting spin stripe order  apparently gaps the spectrum of  Bogoliubov quasiparticles.  Also interesting is the finding by  \textcite{zelli-2011,zelli-2012}  that the pockets of Bogoliubov quasiparticles of the PDW state can give rise to quantum oscillations of the magnetization in the mixed state of a PDW superconductor.

\subsubsection{Amperian Pairing and the PDW state}
\label{amperian}

A recent mean-field theory by \textcite{lee-2014}, based on the concept of Amperian pairing \cite{sslee-2007},  finds a PDW state with pockets of Bogoliubov quasiparticles. In this RVB approach \cite{Baskaran-1987,lee-2006}, the electron is expressed as a composite of  a spinless charged boson (the holon) and a charge neutral spin-1/2 fermion (the spinon).  A necessary accompaniment to this decomposition is that  the holons and spinons are coupled through a strongly fluctuating gauge field, typically with a $U(1)$ gauge group.  In the simplest version, known as the Baskaran-Zou-Anderson (BZA) mean-field state \cite{Baskaran-1987}, this theory assumes that the spinons form a Fermi surface while  the holons Bose condense. In this picture, the superconducting state arises once the spinons form pairs with $d$-wave symmetry and themselves condense \cite{Kotliar-1988}.  Amperian pairing refers to   a pairing of  spinons
  triggered by singular forward-scattering interactions mediated by the gauge bosons  of this theory \cite{sslee-2007}.  The variational wave functions used by \textcite{himeda-2002} and by \textcite{raczkowski-2007}  in the context of the $t$-$J$ model, are in fact  Gutzwiller-projected BZA mean-field states generalized to allow for periodically-modulated pairing as in the PDW state. \textcite{lee-2014} argues that the RVB mean-field theory also allows for a 
 spinon-pair condensate with finite momentum tied to the nesting wave vector of the mean-field theory spinon Fermi surface. From the perspective of broken symmetries, this state is identical to the PDW state.   In agreement with what we advocate here, and in our earlier papers \cite{berg-2009b,berg-2009}, in Lee's theory the uniform SC and the PDW are taken to be the dominant orders, while other orders, {\it e.g.}\ charge-stripe order, are subdominant (see Section \ref{explicit-models}). In the resulting PDW state, the original BZA Fermi surface is gapped out by the spinon condensate leaving behind pockets of charge neutral spinons. The emerging picture has qualitatively the same  features as found  in the Bogoliubov-de Gennes mean-field theory \cite{baruch-2008,berg-2009b}.  Specifically, \textcite{lee-2014} argues that this explains puzzling features of the observed ARPES spectra in 
 {\BSCCO} \cite{he11a}.

\subsubsection{Thermal stabilization of the PDW phase}
\label{thermal}

The variational treatment of the  $t$-$J$ model discussed in Sec.~\ref{variational-t-J} 
shows that the ground-state energy of the PDW is very close to the true ground-state energy;  no convincing evidence has been adduced that it is the true ground state.  However, the PDW has larger low temperature entropy than any of the competing phases \cite{lee-2014}.  Specifically, while any of the phases with a uniform $d$-wave component of the order parameter has, at most, nodal points in $k$-space at which there are gapless quasiparticle excitations, the PDW supports Fermi pockets with a non-zero density of states.  
This gives rise to a contribution to the low-temperature entropy of the PDW of the form $S_{\rm PDW} =\alpha_{S}x k_{\rm B} T/\sqrt{\Delta_0 t} +{\cal O}(T^2)$, where $\alpha_S$ is a number of order 1 and $\Delta_0$ is the scale of the  antinodal gap;   all other states have entropies that vanish at least in proportion to $T^2$.   Even if   the PDW phase has slightly higher energy density than the true ground-state phase by an amount  $\Delta E \approx \alpha_E x t$, then for small enough $\alpha_E$, the PDW will nonetheless be the equilibrium phase for a range of $T$ above $T_- \approx \Delta_0 \sqrt{ \alpha_E/\alpha_S} [t/\Delta_0]^{3/4} \sim 0.03 [t/\Delta_0]^{3/4}$.  It is easy to envisage circumstances (especially where $\Delta_0/t$ is not too small), for which this transition occurs at temperatures above but comparable to the uniform $T_{\rm SC}$.

 \subsection{Superconductivity and Nematic Order}
 \label{nematicity}
 
Charge nematic order has been seen in transport experiments in very underdoped samples of {\YBCO} \cite{ando-2002}, in inelastic neutron scattering on underdoped YBCO in the regime where there is no spin gap \cite{hinkov-2007}, and above $T_{\rm SC}$ for a broad range of hole doping in  Nernst effect measurements \cite{daou10}. STM experiments have documented nematic order on long length scales in underdoped {\BSCCO} \cite{lawler-2010} and have provided evidence for a nematic quantum critical point near optimal doping in {\BSCCO} \cite{fuji14a}, although this interpretation has been challenged  \cite{dasi13}. Charge nematic order was conjectured by two of us in 1998   \cite{kivelson-1998}, including the possible occurrence of such a nematic QCP inside the high $T_c$ superconducting phase. For a recent review of electronic nematic order see \textcite{Fradkin-2010}.
 
However, the relation between charge nematic order and {\hty} has remained unclear.  Part of the problem is that most theories of nematicity have been formulated either in terms of a Pomeranchuk instability in a Fermi liquid \cite{oganesyan-2001} or in terms of the coupling of a Fermi liquid to a somehow preexisting nematic-order-parameter field theory \cite{metlitski-2010},  whereas the nematic state observed in experiments on {\hts} is best described as  a system of ``fluctuating stripes'' \cite{kivelson-2003}, {\it i.e.}\ as a state in which stripe charge order has been melted (thermally or quantum mechanically) leaving a uniform nematic phase \cite{kivelson-1998}.  Nevertheless, several theories have been proposed which suggest that, inside the nematic phase, quantum nematic fluctuations may give rise to a SC state \cite{kee-2004}, and that superconductivity which is primarily the result of other pairing interactions ({\it e.g.}, spin fluctuations) can be enhanced in the neighborhood of a nematic QCP \cite{metlitski-2014,lederer-2014}. The relation (if any) between nematicity and PDW  order is even less clear. However,  it has been shown recently  that near a QCP of an (as-yet undectected)  spin-triplet nematic state \cite{wu-2007} a host of SC states arise, including  PDW phases \cite{soto_garrido-2014}.

\section{Intertwined order in the cuprates}
\label{experiments}

Here we review experimental evidence for many of the types of order that have already been 
discussed.  In particular, we emphasize the close associations between various distinct orders.

\subsection{SDW, SC, and PDW orders}

Perhaps the most striking %connection involves
%\textcolor{green}{
evidence of an intimate connection between   SDW and SC orders is observed  in the ``214'' cuprates derived from La$_2$CuO$_4$.  For example, in La$_2$CuO$_{4+\delta}$ with stage-4 interstitial order \cite{lee04a} SDW and SC orders both onset at 40~K in zero magnetic field.  While application of a $c$-axis magnetic field causes a reduction in $T_{\rm SC}$, it changes $T_{\rm SDW}$ relatively little, although it enhances the SDW order.  The story is similar in \LSCO\ with $x=0.10$,  where application of a $c$-axis magnetic field induces SDW order that onsets %essentially 
at the zero-field $T_{\rm SC}$ \cite{lake02}. 

\begin{figure}[tb]
\begin{center}
\includegraphics[width=0.3\textwidth]{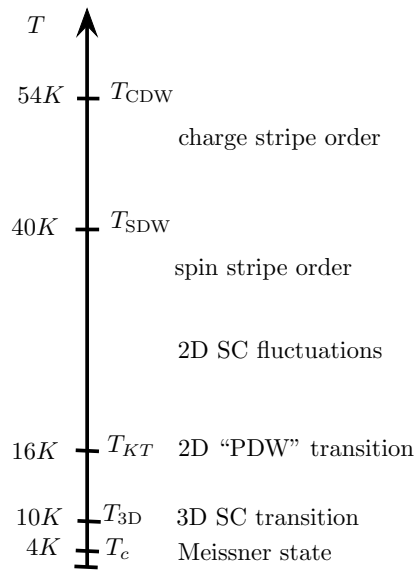}
\end{center}
\caption{Scales and phases of {\LBCO} near $x=1/8$: PDW fluctuations begin below the onset of charge order $T_{\rm CDW}$, and become pronounced below $T_{\rm SDW}$. The (presumably) PDW 2D SC phase below $T_{\rm KT}$. The regime between $T_{\rm 3D}$ and $T_c$ is expected to be an $XY$ glass, and the 3D d-wave SC lies below $T_c$. See text for details.}
\label{fig:scales-LBCO}
\end{figure}

%EF
%\textcolor{blue}{
%A case of particular interest is 
The most compelling evidence of PDW order is found in {\LBCO}. %, which is the original high $T_c$ superconductor \cite{bedn86}. 
 %The phase diagram of this material shows  two SC lobes (or domes), each with a $T_c \sim 38$K, instead of a single dome, with the minimum $T_c$ found near doping $x=1/8$ (the `1/8 anomaly'). 
 Unlike the typical cuprate, the highest $T_c \approx 32$K occurs at $x \approx 0.09$, and there is  a deep minimum in the $T_c$ vs.\ $x$ curve at  $x=\frac18$ [the `1/8 anomaly'; see, {\it e.g.},\citet{huck11}]. 
However, even %in \LBCO\ with 
at $x=\frac18$, %where the bulk $T_{\rm SC}$ is strongly depressed, 
quasi-2D SC correlations onset %together with SDW order near 40 K 
above 40~K \cite{li07,tran08}; %here, the SC correlations have been attributed to PDW order, 
the existence of such a 2D regime  %which provides an explanation
can be most readily understood as being a consequence of % for the evident
the frustration of interlayer Josephson coupling \cite{berg-2007,berg-2009b,himeda-2002} associated with PDW order. Indeed, at $x=\frac18$ \LBCO\ exhibits a cascade of phase transitions and crossovers, as shown in Fig.\ref{fig:scales-LBCO}. In order of descending temperatures, this system has a charge order transition at $T_{\rm CDW}=54$~K where charge stripe (CDW) order onsets (roughly coinciding %in this material 
with a structural transition). At $T_{\rm SDW}=42$~K long range SDW order %has its
  onsets. 
 As a function of decreasing $T$, $\rho_{ab}$ drops by an order of magnitude just below  $T_{\rm SDW}$, then levels off before again dropping precipitously to immeasurably small values as $T$ approaches  $T_{\rm KT}=16$~K.  Meanwhile, $\rho_c$ begins to increase significantly below $T_{\rm CDW}$, reaching a maximum value at around 30~K, and finally dropping precipitously toward zero 
  %In between these two temperatures transport data indicates that the c-axis resistivity grows with lowering temperatures while the in-plane resistivity $\rho_{ab}$ is essentially constant. Below $T_{\rm SDW}$ $\rho_c$ continues to rise with decreasing temperatures while $\rho_{ab}$ becomes very small quite rapidly. At a temperature $T_{KT}=16$K the in-plane resistivity $\rho_{ab}$ vanishes while $\rho_c$ is still quite robust.  Below $T_{KT}$ LBCO behaves as a layer 2D superconductor without a measurable c-axis Josephson coupling, i.e. the layers are dynamically decoupled. The 3D resistive transition is only reached at a lower temperature,
  as $T$ approaches $T_{\rm 3D}=10$~K.  For $T_{\rm KT}> T > T_{\rm 3D}$, LBCO behaves as if  the layers were decoupled, leaving a 2D superconductor without detectable c-axis Josephson coupling.  Moreover, the %3D %SC 
full (3D)  Meissner state is only observed below $T_c=4$~K. 
%EF

The first evidence for layer decoupling actually came from a study of $c$-axis optical conductivity in La$_{1.85-y}$Nd$_y$Sr$_{0.15}$CuO$_4$ \cite{taji01}, where it was observed that the Josephson plasma resonance (JPR) essentially disappeared when the Nd concentration was tuned into the stripe-ordered regime.  It was later demonstrated that application of a $c$-axis magnetic field to superconducting La$_{1.9}$Sr$_{0.1}$CuO$_4$ causes a rapid reduction in the JPR frequency \cite{scha10}, while resulting in relatively little change in the in-plane superfluid density \cite{scha10b}.

Further evidence for a close association between CDW/SDW and SC order comes from pump-probe studies.  La$_{1.8-x}$Eu$_{0.2}$Sr$_x$CuO$_4$ has the same low-temperature structure as \LBCO, exhibits CDW and SDW order, but has more strongly suppressed SC order \cite{fink11}.  Nevertheless, it has been demonstrated that pumping a crystal of $x=\frac18$ doping with a very short burst of  80-meV photons can induce the appearance of SC order, as indicated by probing the $c$-axis infrared reflectivity, in a time as short as 1 to 2 ps after the pump \cite{faus11}.    In a related experiment \cite{fors14} on \LBCO\ with $x=\frac18$, it has now been shown that the pump causes the CDW order to melt within $\sim0.4$ ps; the crystal symmetry shows a slight response, but only on a delayed timescale of 15 ps.  The fast turn on of the bulk superconductivity in a sample with a stripe-ordered ground state suggests that strong superconducting correlations are also present in that ground state.  The fact that the same pump melts the charge stripes is compatible with the idea that PDW order in the ground state frustrates the interlayer superconducting phase coherence.  Disrupting the static CDW order also 
affects the PDW, removing the frustration and resulting in bulk superconductivity.

For the pump to be effective at inducing interlayer coherence in the $x=0.125$ samples, the polarization must be parallel to the CuO$_2$ planes, within which it couples to Cu-O bond-stretching phonons.  Intriguingly, a new experiment on \LBCO\ with $x=0.115$ demonstrates that a near-infrared pump pulse with polarization perpendicular to the planes can induce a $c$-axis JPR at a temperature as high as 45~K, well above the bulk $T_{\rm SC}$ of 13 K, but below $T_{\rm CDW}$ \cite{nico14}.

Continuing with \LBCO\ but moving to $x=0.095$, bulk SC onsets together with weak SDW order at 32 K \cite{wen12}.  In this case, weak CDW order appears at the same temperature, constrained by the structural transition \cite{wen12a}.  Enhancement of SDW and CDW order by an applied magnetic field occurs at the expense of the SC order \cite{wen12,huck13}; nevertheless, the coincident onset temperatures indicate a close connection between these orders.  In \LNSCO, there is evidence for a similar onset of quasi-2D superconductivity at a temperature well above the bulk $T_{\rm SC}$ \cite{ding08}; however, the onset of SDW order occurs at a temperature $\sim20$ K higher \cite{ichi00}.

An interesting situation occurs in electrochemically oxygen-doped La$_{2-x}$Sr$_x$CuO$_{4+\delta}$ \cite{moho06}.  Here, for $x\lesssim\frac18$, the excess oxygen content appears to tune itself so that the net hole concentration is approximately $\frac18$.  Measurements by muon spin rotation ($\mu$SR) spectroscopy indicate that the volume is phase separated into superconducting and magnetically-ordered regions \cite{moho06}.  Nevertheless, neutron diffraction measurements on crystals with $x=0.04$, 0.065, and 0.09 find that SDW peaks appear simultaneously with the SC order at $\sim 39$~K for all samples \cite{udby13}.  Regardless of whether these orders are entirely segregated, their energy scales are remarkably similar.

Evidence of anomalous 2D superconducting fluctuations in \YBCO\ exist, although their association with PDW order is far less clear than in {\LBCO}.  Optical evidence of a  JPR within a bilayer persisting to temperatures well above $T_{\rm SC}$ has been presented by \textcite{dubr11}.  Interestingly, there seems to be some correlation between the onset temperature of the JPR and the onset of CDW and/or nematic order.  Recent magnetization studies in high fields by \textcite{yu14} have likewise been interpreted as evidence of significant PDW correlations.  Finally, pump-probe studies similar to those reported in \LBCO\ have been carried out in \YBCO, with results that, while less clear-cut in terms of magnitude and persistence, are still reminiscent of the former \cite{hu14,kais14}.  It is a key issue to determine if PDW-type correlations exist in YBCO and more generally in hole-doped cuprates.  Interestingly, the latest experiment \cite{fors14} on underdoped YBa$_2$Cu$_3$O$_{6.6}$ indicates that the pump conditions that enhance the coherent interlayer transport at $T> T_{\rm SC}$ also depress the CDW order of the type that will be discussed in Sec.~\ref{exp:cdw_sc}.

\subsection{CDW and SDW orders}

For a number of 214 compounds, CDW order develops at a temperature that is generally higher than the SDW transition.  For \LBCO\ and \LNSCO, the CDW order is limited by a structural transition that breaks the effective 4-fold symmetry of the Cu-O bonds \cite{axe94,huck11,ichi00}; however, in La$_{1.8-x}$Eu$_{0.2}$Sr$_x$CuO$_4$, where the structural transition takes place at $T>120$~K, the maximum $T_{\rm CDW}$ is a modest 80~K
 \cite{fink11}.  The fact that $T_{\rm CDW} > T_{\rm SDW}$ indicates that the CDW order is not secondary to the SDW \cite{zachar-1998}, in contrast to the situation in chromium \cite{pynn76}.  This does not mean that the SDW and CDW are not strongly correlated with one another---neutron-scattering experiments show that the SDW fluctuations become virtually gapless as soon as CDW order is established \cite{fuji04,tran08}.

In \LSCO, where the average crystal structure makes all Cu-O bonds equivalent, relatively strong SDW order is observed only for $x\approx0.12$
 \cite{yama98,kimu00}.  Nevertheless, a nuclear quadrupole resonance (NQR) study  detected a pattern of intensity loss in the normal state of underdoped \LSCO\ that matches the behavior observed in CDW-ordered 214 cuprates \cite{hunt99}.  While the direct cause of intensity loss is likely from SDW correlations \cite{juli01}, CDW order has now been detected by x-ray diffraction for $x$ near 0.12 and $T< 85$~K \cite{wu12,chri14,tham14,crof14}. Interestingly, both the SDW and CDW wave vectors are rotated $\sim3^\circ$ from the Cu-O bond directions \cite{kimu00,tham14}.  
From a symmetry  perspective \cite{robertson-2006}, this rotation is a necessary consequence of the incommensurate character of the SDW and the orthorhombicity of {\LSCO}.  However, the magnitude of the effect  surely reflects the fact that diagonal SDW order dominates for $x<0.055$ where
 {\LSCO\} is insulating \cite{birg06}.\footnote{Recall (Sec. \ref{variational-t-J}) that in variational  studies of the $t-J$ model, insulating diagonal stripes were found to have energy only slightly larger than that of vertical stripes, and so could be easily stabilized in an orthorhombic environment.} For $T<T_{\rm SC}$, application of a $c$-axis magnetic field enhances both the SDW and CDW order, 
in much the same way is in LBCO for dopings sufficiently far from $x=\frac18$ that  stripe order is weak in zero field \cite{wen12,huck13}.

A significant feature of the CDW order in 214 cuprates is that the wave vector is locked to that of the SDW, with 
${\bm q}_{\rm CDW}=2{\bm  q}_{\rm SDW}$.  To distinguish this feature, we will use the label CDW1 to denote it in comparisons below.

\subsection{CDW/SDW quantum critical point}

Several studies have 
documented possible signatures of quantum critical behavior near the doping 
at which CDW order disappears in 214 systems \cite{doir12}.  In La$_{1.8-x}$Eu$_{0.2}$Sr$_x$CuO$_4$, $\mu$SR studies have detected SDW order over a broad range of $x$, ending near $x_c\gtrsim0.2$ \cite{klau00}.  Oxygen-isotope effect \cite{sury05}, Hall effect \cite{take04}, and resonant soft-x-ray diffraction \cite{fink11} studies suggest that CDW order disappears at a similar $x_c$.  Superconductivity is observed for $0.14 < x < 0.27$, with $T_{\rm SC}$ forming a dome centered on $x\sim0.21$ \cite{sury05}.  This looks similar to cases of SC order appearing at a quantum critical point, with the major difference that the magnitude of $T_{\rm SC}$ (20~K) is comparable to the maxima of $T_{\rm SDW} \approx 27$~K and $T_{\rm CDW}\approx 80$~K.

In \LNSCO, neutron and x-ray diffraction measurements suggest that there should be a similar $x_c$ where stripe order disappears \cite{ichi00,niem99}, and the maximum of $T_{\rm SC}$ occurs at a similar position \cite{daou09b}.  Measurements of the in-plane resistivity in a magnetic field sufficient to suppress the superconductivity indicate linear-$T$ behavior for $x=0.24$ down to low temperature, but an upturn below 40 K for $x=0.20$, with a related difference in the Hall effect \cite{daou09b}.  These behaviors, together with results on the thermopower, are consistent with the presence of a quantum critical point  at $x=x_c$ \cite{daou09a}.

\subsection{CDW and nematic orders}

Spectroscopic imaging scanning tunneling microscopy (SI-STM) has provided evidence for short-range CDW correlations in \BSCCO\ \cite{hoff02,howa03b,kohs07,park10,dasi14,fuji14a},  Bi$_{2-y}$Pb$_y$Sr$_{2-z}$La$_z$CuO$_{6+x}$ \cite{wise08}, and \oxychloride\  \cite{kohs07}.  Besides having a short correlation length, another difference from the 214 cuprates is that the CDW wave vector decreases with doping \cite{wise08,dasi14}, scaling roughly like the antinodal $2k_{\rm F}$ measured by angle-resolved photoemission spectroscopy but larger in magnitude \cite{meng11,comi14}.  Because of this difference, we will denote the order as CDW2.

While while the translational-symmetry breaking of the charge is local, a long-range rotational symmetry breaking associated with an electronic nematic state \cite{kivelson-1998} has been identified at low temperature in underdoped \BSCCO\   \cite{lawler-2010,mesa11}.  [Note that this identification is not without controversy \cite{dasi13}.]  With doping, the nematic order and CDW2 correlations both disappear at $x_c \approx 0.19$, the point at which the low-temperature antinodal pseudogap appears to close \cite{tall01,vish12,gorkov-2006,fuji14a}.  Studies of quantum oscillations in \YBCO\ find a mass divergence very nearby, at $x_c \approx 0.18$, suggesting a QCP \cite{rams14}.  Notably, $T_{\rm SC}$ 
 and $H_{c2}$ are maximized here.

While static AF order is not relevant to this behavior, there is an interesting connection with the energy range in which dynamic AF correlations remain strong.  In the parent AF insulator phase, the spin waves are well defined \cite{head10} because they exist at energies ($\lesssim0.3$~eV) far below the gap for charge excitations [$\sim 1.5$~eV \cite{baso05}].   This situation changes with hole doping.  Experimentally it has been observed that the momentum-integrated magnetic spectral weight remains comparable to that of the parent insulator for energies below the antinodal pseudogap energy, becoming much weaker above that scale \cite{stoc10,fuji12a}.  With doping, the magnetic spectral weight close to the AF wave vector becomes quite weak for $x\gtrsim0.2$ \cite{fuji12a}.  Thus, there is at least a strong association between nematic order and AF spectral weight, consistent with the idea that
related electronic textures are necessary to sustain even short-range AF correlations for substantial $x$.

To get a measure of the temperature scale associated with nematic order, we must turn to \YBCO, where a study of in-plane anisotropy in the Nernst effect has suggested that nematic order develops at a temperature comparable with $T^\ast$ determined from in-plane resistivity \cite{daou10}.  Of course, in orthorhombic \YBCO, the 4-fold symmetry of the planes is already broken by the presence of the Cu-O chains; nevertheless, the temperature-dependence of the anisotropy in the Nernst effect is quite distinct from that of the orthorhombic strain.  The relatively sharp onset \cite{xia08,he11a,kara12} of a Kerr signal  in multiple families of cuprates at similar temperatures to those at which charge order begins to be detectable is probably associated \cite{hosur-2013,hosur-2014,varma-2013} with some pattern of point-group symmetry breaking, as well.

At the onset of the anisotropic behavior, the Nernst coefficient is found to be negative \cite{daou10}.  As the temperature drops and approaches $T_{\rm SC}$, a positive contribution to the Nernst coefficient develops that is associated with superconducting fluctuations \cite{wang06}.  In 214 cuprates, a positive contribution to the Nernst coefficient is also detected in association with CDW order \cite{cyrc09,hess10,li11}, although the magnitude of this contribution is depressed for $x\approx\frac18$.  The trend is somewhat different in \YBCO\ with $x=0.12$ \cite{chan11}, where suppression of the superconductivity with a strong magnetic field leaves the Nernst coefficient strongly negative.  This occurs in the regime where various measures of CDW order have been reported \cite{wu11,ghir12,chan12a}, as we discuss next.

\subsection{CDW and SC orders}
\label{exp:cdw_sc}

In \YBCO, at least two CDW phase boundaries have been detected.  Starting with the case of zero magnetic field, short-range CDW order onsets gradually below a transition temperature that has a maximum of $\sim150$~K for $x\sim0.12$ \cite{ghir12,chan12a,achk12}.   The onset temperature and strength of the order both decrease as doping approaches the regime of quasi-static magnetic order at $x\lesssim0.08$ and the regime of optimum doping at $x\gtrsim0.14$ \cite{ghir12,blac13a,huck14,blan14}.  The intensity grows on approaching $T_{\rm SC}$ from above, and then decreases somewhat on cooling below $T_{\rm SC}$.  Application of a $c$-axis magnetic field enhances the CDW intensity for $T<T_{\rm SC}$, but has no impact for $T>T_{\rm SC}$
 \cite{chan12a,blac13a}.  Initially, there were some questions as to whether the response detected by x-ray scattering might be dynamic, with integration over the fluctuations by coarse energy resolution; however, recent characterizations of the inelastic spectrum \cite{blac13b,leta14} and the comparative strength of the scattering \cite{tham13} indicate that the CDW correlations are static.  The fact that a broadening of the NQR line is seen \cite{wu14} with similar $T$ dependence as the X-ray  signal, further corroborates the static character of the CDW correlations.

The general doping dependence of the CDW order, with a maximum at $x\sim0.12$, and the enhancement by a magnetic field that also suppresses SC order, are similar to features observed in 214 cuprates.  In contrast, there is a significant difference in the CDW wave vectors.  First of all, there are distinct modulations along the principal axes, ${\bf q}_1 = (\delta_1,0,0.5)$ and ${\bf q}_2 = (0,\delta_2,0.5)$, with $\delta_2$ slightly greater then $\delta_1$ and both having a magnitude close to 0.3 \cite{ghir12,chan12a,blac13a}.  While the intensities of the CDW peaks at ${\bf q}_1$ and ${\bf q}_2$ can be comparable for a range of $x$, they can also be quite different, as in the phase with Cu-O chain order characterized as ortho-II \cite{blac13a,blan13}.

The second difference from 214 cuprates has to do with the doping dependence of the CDW wave vectors, which decrease with $x$.  This behavior is similar to that seen by STM in, for example, \BSCCO, and, in fact, x-ray scattering experiments have demonstrated CDW peaks in \BSCCO\
 \cite{dasi14,hash14} and Bi$_2$Sr$_{2-x}$La$_x$CuO$_{6+\delta}$ \cite{comi14} at wave vectors identical to those inferred from STM measurements on the same samples.  As a result, this order corresponds to CDW2.  Similar CDW2 order has also been detected in underdoped HgBa$_2$CuO$_{6+\delta}$ by x-ray scattering \cite{tabi14}.

Incommensurate AF correlations have a finite excitation gap across most of the doping range where CDW order is observed \cite{dai01,hink10}.  In fact, the spin gap appears to open on cooling at essentially the same temperature as $T_{\rm CDW}$, based on NMR measurements of the spin-lattice relaxation rate, $1/T_1$ \cite{huck14,baek12}.  The wave vector of the lowest-energy AF correlations grows with $x$ in a manner qualitatively similar to that seen in both \LSCO\ and Bi$_{2-x}$Sr$_{2+x}$CuO$_{6+\delta}$ \cite{enok13}.  Substitution of Zn into planar Cu sites in \YBCO\ induces local, static, short-range SDW correlations while suppressing the CDW2 intensity \cite{blan13}.   Also, for \YBCO\ with $x\lesssim0.08$, where CDW2 order fades away, the spin gap collapses, and the low-energy spin correlations have uniaxially-modulated
short-ranged SDW character \cite{haug10}.

Considering the similar doping ranges for CDW1 and CDW2, together with their opposite relationships with SDW order, as well as the universal presence of dynamic AF correlations (of correlated-insulator character) across the underdoped regime, it appears as if CDW1 and CDW2 are dual characters of the same underlying electronic texture, possibly nematic.  
However, only one of these aspects is realized at a time. 

The current interest in CDW order was originally stimulated by observations of quantum oscillations in various transport properties measured in \YBCO\ as a function of magnetic field \cite{doir07,lebo07,seba08}.  The low observed oscillation frequency implies small pockets, presumably due to Fermi-surface reconstruction \cite{tail09,seba12}, stimulating analyses in terms of stripe \cite{millis-2007} and CDW order \cite{yao-2011}; such a connection continues to influence the interpretation of quantum oscillation studies \cite{seba14}.  The temperature dependence of transport properties measured in high magnetic fields strongly resemble the low-field behavior of stripe-ordered systems \cite{lali11}.  Direct evidence for CDW order (CDW3, distinct from CDW2)  at high magnetic fields and low temperature has been obtained through NMR measurements \cite{wu11,wu13a}. In the CDW2 phase seen by x-ray scattering, NMR measurements detect broadening of NMR lines from planar O and Cu sites \cite{wu14}, whereas, in the CDW3 phase, a line splitting develops above an onset field of at least 10~T.   The onset temperature for CDW3 is comparable to the zero-field $T_{\rm SC}$.  Measurements of sound velocity \cite{lebo13} suggest that there is a thermodynamic phase transition associated with CDW3; however, the transition fields determined by the sound velocity and NMR studies differ by a significant amount (17 T vs.\ 10 T, respectively).  The relationship of CDW3 to CDW1 and CDW2 remains to be determined; however, the spin correlations remain gapped in CDW3, except near $x=0.08$ \cite{wu11,wu13b}.

\section{Superconducting vs. Pseudo gaps} 
\label{Pseudogap}

No issue in cuprate physics, it seems, has been more intensely debated than the nature and origin of the pseudogap, $\Delta_{\rm pg}$.  Many of the differing perspectives have been reviewed by others \cite{timu99,tall01,norman-2005,hufn08}.  Here we wish to summarize some of the empirical observations regarding superconducting and pseudogap behavior in the cuprates, and then discuss how the concept of intertwined orders  
might provide a consistent interpretation.

\subsection{Spectroscopic characterizations}

Empirically, two characteristic energy gaps have been identified in the cuprates \cite{deut99,hufn08}. One of these, $\Delta_{\rm c}$, is a measure of the coherent superconducting gap, as obtained, for example, from Andreev reflection in point-contact spectroscopy with current flowing along an in-plane Cu-O bond direction.  As a function of doping, $\Delta_{\rm c}$ forms a dome, with $2\Delta_{\rm c}/k_{\rm B}T_{\rm SC}\sim4$--6 \cite{deut99}, as indicated schematically in Fig.~\ref{fg:gap_v_p}(top).   The other gap, the pseudogap $\Delta_{\rm pg}$, can be measured by tunneling along the $c$ axis.  In the over-doped regime, $\Delta_{\rm pg}\sim\Delta_{\rm c}$, but as $x$ decreases, $\Delta_{\rm pg}$ grows monotonically \cite{deut99,hufn08}.  $\Delta_{\rm c}$ is only detected for $T<T_{\rm SC}$, whereas $\Delta_{\rm pg}$ can be observed up to $T\sim T^*$ \cite{timu99,hufn08}, as indicated in Fig.~\ref{fg:gap_v_p}(bottom). 

\begin{figure}[bt]
\centerline{\includegraphics[width=0.8\columnwidth]{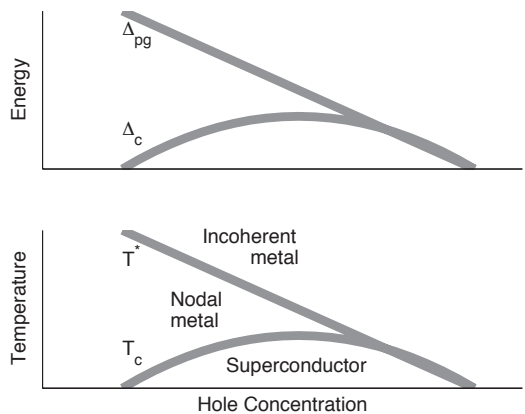}}
\caption{(color online)
Schematic diagram summarizing doping dependence of gaps and characteristic temperatures.  Above optimal doping, the experimental identification of the pseudogap can differ depending on whether one is looking at measurements in the normal or superconducting state; the behavior shown is consistent temperature-dependent tunneling measurements \cite{deut99}.
}
\label{fg:gap_v_p} 
\end{figure}

Further information on the gaps and their relationship is provided by angle-resolved photoemission spectroscopy (ARPES) \cite{dama03,hash14b}, much of which has been done on \BSCCO, because of its excellent cleavability.  Spectroscopic imaging with scanning tunneling microscopy (STM) provides further information \cite{fisc07,fuji12c}.  Where ARPES averages over a substantial surface area, STM is able to map the microscopic variation of states.

With the possible exception of samples with $x \lesssim 0.08$, ARPES experiments on {\BSCCO} deep in the superconducting state ($T \ll T_{\rm SC}$) exhibit a Fermi surface gap with d-wave symmetry.  For nearly optimal doping, the gap has the simple angular dependence, $\Delta({\bf k}) \approx  \Delta_0[\cos(k_x)-\cos(k_y)]$, as indicated in Fig.~\ref{fg:fs_gaps}.  While the near-nodal gap appears to be approximately $x$ independent for a  broad range of doping below optimal, \cite{vish12,dasi13}  the antinodal gap increases  with decreasing $x$.    Sharp  ``quasiparticle'' peaks with energy 
$\Delta({\bf k})$ and width small compared to the  antinodal gap energy exist along the entire Fermi surface, again with the possible exception of highly underdoped samples. \cite{vish12,zhao13}.
Likewise, in single-layer (Bi,Pb)$_2$(Sr,La)$_2$CuO$_{6+\delta}$, quasiparticle peaks are harder to identify, but by looking at the difference between the spectra above and below $T_c$, coherent quasiparticle-like features have  been identified  around the entire Fermi surface, but only  for doping  above optimal \cite{kond09}.   However, for $x\lesssim 0.18$ a rather different picture emerges from an analysis of the  quasiparticle interference (QPI) signal measured in STM.  From this it is inferred that there are no coherent quasiparticles in the antinodal regime (beyond the AF Brillouin zone boundary) and moreover, that there is effectively a nodal arc---despite the fact that one is deep in the superconducting state \cite{he14,fuji14a}.

For underdoped and even slightly overdoped samples, although the gap in  the antinodal portion of the Fermi surface unambiguously persists  \cite{loes96,ding96,kani06} for a range of temperatures  above $T_{\rm SC}$ and below $T^*$, simple measures indicate that the gap closes at $T_{\rm SC}$ for along a finite ``Fermi arc'' \cite{norm98} in the near nodal region, as indicated in Fig.~\ref{fg:fs_gaps}.  The  antinodal $\Delta_0$ thus corresponds to $\Delta_{\rm pg}$, while the coherent gap $\Delta_c$ seemingly corresponds to $\Delta(k)$ at the wave vectors corresponding to the ends of the normal-state Fermi arc \cite{push09,rame11,rebe13}. However, the arc ends are probably not very well defined; indeed, it is unclear whether the arcs are produced by the vanishing of the nodal gap \cite{lee07}, or simply indicate the portion of the Fermi surface in which the  scattering rate is greater than the gap \cite{kond13,rebe13}.  The near-nodal scattering rate appears to be strongly $T$ dependent near $T_c$ in this analysis.
Indeed, while the energy of the antinodal quasiparticles does not change to any detectable extent upon approach to $T_c$, the coherent spectral weight ({\it i.e.}, the peak in the spectral function at $\Delta({\bf k})$) vanishes at $T_c$ or slightly above \cite{fedo99}.

As mentioned above, there are significant differences in aspects of the quasiparticle spectrum measured in ARPES and those  inferred from the QPI analysis of the STM spectrum.  Reconciling these -- which we will not attempt here -- is a major open issue.  
Still, in the superconducting state, STM shows that, in real space, the gap behavior is spatially uniform for $E<\Delta_c$, but inhomogeneous at higher energies, especially near $\Delta_{\rm pg}$ \cite{howald-2001,pan01,push09}.  

\begin{figure}[t]
\centerline{\includegraphics[width=0.9\columnwidth]{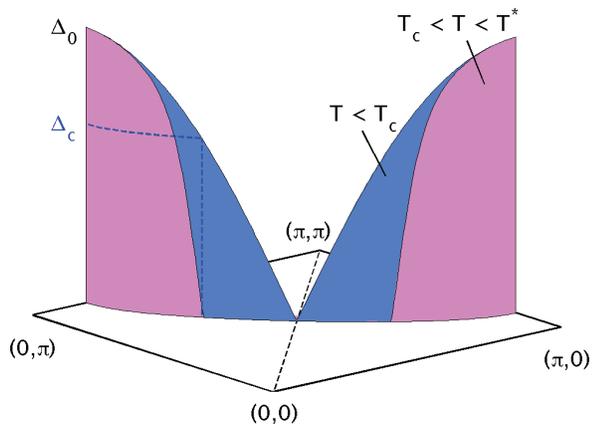}}
\caption{(color online)
Schematic diagram summarizing ARPES measurements of gaps around the Fermi surface in underdoped cuprates. The combined gray (blue and magenta) regions indicate the $d$-wave gap, with magnitude $\Delta_0$, observed at $T<T_{\rm SC}$.  At $T_{\rm SC}$, the dark gray (blue) part of the gap closes, leaving a gapless arc of states in the near-nodal region and a pseudogap (light gray/magenta) in the antinodal region.  The pseudogap loses definition at $T\sim T^*$.  The energy scale for coherent pairing, $\Delta_c$, indicated by tunneling empirically corresponds to the magnitude of the $d$-wave gap at the wave vector corresponding to the end of the nodal arc. 
}
\label{fg:fs_gaps} 
\end{figure}

\subsection{Temperature dependence and broader context}

The temperature dependence of the pseudogap can be somewhat ambiguous, so it is instructive to consider results from a variety of experimental probes.  For example, $c$-axis optical conductivity provides a useful probe of $\Delta_{\rm pg}$, as the contribution from quasiparticles in the nodal region is suppressed.  The temperature dependence of $\sigma_c(\omega)$ in underdoped \YBCO\ suggests that $\Delta_{\rm pg}$ does not change but rather the
 states gradually fill in as the temperature approaches $T^*$ \cite{home93b}.  A similar impression is given by an analysis of the temperature dependence of the Hall constant in \LSCO\ \cite{gorkov-2006}.  Fitting with a temperature-independent term (from near-nodal carriers) plus a thermally activated component yields a temperature-independent gap that is quantitatively consistent with 
 $\Delta_{\rm pg}$ determined by ARPES.   
 STM measurements indicate that the pseudogap is detectable (with a spatially varying magnitude) in the entire field of view for a significant range of $T>T_{\rm SC}$, but becomes   confined to increasingly rare regions as the temperature approaches $T^*$ \cite{gome07}.  
In ARPES, the  pseudogap onsetting at $T^*$ 
has long been associated with the antinodal region. (For recent references, see \onlinecite{kond13,he11a}.)

Most of the spectroscopic measures of the pseudogap contain no direct information on the origin of the gap.  There are a few measurements that provide evidence for pairing correlations.  One of the key features of superconducting pairs is that they involve a mixing of particle and hole states.  This would result in particle-hole symmetry near the Fermi energy.    
 ARPES measurements of \cite{yang08,yang11} on underdoped \BSCCO\ found evidence of such particle-hole symmetry for antinodal states far above $T_{\rm SC}$.   The mixing of particle and hole states also impacts the dispersion of the particle states, and this characteristic dispersion has been seen for antinodal states above $T_{\rm SC}$ in the same system \cite{kani08}.  On the other hand, ARPES studies of single-layer Bi$_{1.5}$Pb$_{0.55}$Sr$_{1.6}$La$_{0.4}$CuO$_{6+\delta}$ \cite{hash10} find significant particle-hole asymmetry in similar analysis.  In underdoped \YBCO, a different signature of superconducting correlations in the normal state has been reported.  Measurements of optical conductivity for light polarized along the $c$ axis find indications of a transverse Josephson plasma resonance at temperatures as high as 180~K \cite{dubr11},  and enhanced coherent transport at $T\gg T_{\rm SC}$ has been achieved by optically pumping apical oxygen vibrations \cite{hu14}. 

It is also relevant to note that there is another significant gap in the problem.
The undoped cuprate parent compounds are charge-transfer insulators, with an optical gap ($\sim 1.5$~eV) limited by the energy difference between Cu $3d_{x^2-y^2}$ and O $2p_\sigma$ states rather than the larger onsite Coulomb repulsion $U$ that is responsible for magnetic moments on the Cu sites
 \cite{emery-1987,zaanen-1985}.  On doping holes into the CuO$_2$ planes, the optical gap does not collapse; rather, infrared reflectivity studies demonstrate the coexistence of the charge-transfer gap with finite optical conductivity that is transferred into the gap.  Integrating the conductivity within the gap, the effective carrier density grows
 in proportion to $x$ \cite{coop90,uchi91}, rather than $1-x$ as predicted by conventional band theory.  At high temperatures, the conduction electrons are completely incoherent, as indicated by the absence of a Drude peak in the optical conductivity
  \cite{lee05,take02}.  A Drude peak develops on cooling, with a weight that is proportional to, but smaller than, $x$ within the underdoped regime
  \cite{uchi91,padi05}.   The Drude peak is likely associated with near-nodal states and develops as the pseudogap becomes apparent in the antinodal region; hence, the region $T_{\rm SC}<T<T^*$ is labelled ``nodal metal'' in Fig.~\ref{fg:gap_v_p}(bottom).  Of course, these features are associated with energy scales much smaller than the charge-transfer gap and, especially, $U$.

\subsection{Interpreting the pseudogap}

The extent to which this gap, especially at temperatures above $T_{\rm SC}$, reflects ``$d$-wave superconducting pairing without phase coherence'' vs.\ a distinct ``second gap'' associated with some other ordering phenomenon (or ``Mottness'')  has been endlessly debated. This is commonly referred to in the literature as the ``one gap'' vs. ``two gaps'' dichotomy.
We will add to this debate in Sec.~\ref{discussion}.  Here we note that a
key issue
is
whether there exists a  crossover temperature below which the amplitude of the order parameter (or  parameters) is well defined.  More specifically, can one phenomenologically associate with an experimentally determined pseudogap temperature, $T^*$, in the phase diagram of the cuprates,  a crossover scale below which many of the spectroscopic characteristics of an ordered phase begin to be apparent, but without any associated long-distance correlations, much less a broken symmetry?  In theory,  it is sometimes possible to identify a crossover temperature  $T_{\rm mf} > T_{\rm SC}$ at which a local amplitude of the order parameter develops---often $T_{\rm mf}$ is identified with a mean-field transition temperature calculated in one way or another. 
This is possible, for instance, in an array of weakly Josephson coupled superconducting grains.
However, it is important to realize that in generic problems, {\it neither the notion of a locally defined amplitude of an order parameter nor $T_{\rm mf}$ are well defined even in theory.}

From the studies of underdoped samples, there are certainly features that suggest two distinct gaps.
The near nodal gap seen in ARPES is undoubtedly a superconducting gap, 
given its  singular angle dependence and the fact that it 
closes, in some sense, more or less at $T_{\rm SC}$. 
Conversely, therefore, it is natural to posit that the antinodal gap, which hardly varies as $T$ crosses $T_{\rm SC}$, is not a superconducting gap.  Moreover, the $x$ dependence of  $\Delta_{\rm c}$ contrasts markedly with that of $\Delta_{\rm pg}$.  Finally, from STM measurements, it appears that $\Delta_{\rm pg}$ tends to be largest in local regions in which evidence of CDW order is strongest, and weakest where CDW correlations are weak \cite{kohs08}, inviting an association between the pseudogap and CDW order.

 In contrast,
the argument for a gap with a single origin begins at optimal doping.
The simple $d$-wave form of the gap at $T\ll T_{\rm SC}$, and the fact that even the energy widths of the quasiparticle peaks do not vary substantially as a function of position along the Fermi surface, make a very compelling case that it is a uniform $d$-wave superconducting gap.  However,  in {\BSCCO}, even for $x=x_{\rm opt}$,  the antinodal gap survives to $T >T_{\rm SC}$ without any significant change in magnitude, {\it i.e.} it becomes the pseudogap.  If it is a superconducting gap below $T_{\rm SC}$, and its magnitude remains roughly the same above $T_{\rm SC}$, 
 it is difficult to 
 imagine it is unrelated to superconducting pairing.
 
 Returning to low $T$, where the antinodal gap increases with decreasing $x$, one might be tempted to associate the ``extra'' gap size with the growth of a second order parameter.  Assuming the two gaps add in quadrature, this would mean that $\Delta_{\rm pg} = \sqrt{|\Delta_0|^2 + |\Delta_{\rm other}|^2}$, which  still implies that the largest contribution to the gap comes from superconductivity as long as $\Delta_{\rm pg} < \sqrt{2}\Delta_0$, as is true at all but the smallest values of $x$.   
From a theoretical perspective, in the weak coupling limit the superconducting instability would be strongly suppressed and even the dominant pairing symmetry would be changed were $\Delta_{\rm pg}$   entirely associated with a partial gapping of the Fermi surface produced by a non-superconducting order \cite{cho-2013,mishra-2014}.

In a loose sense, an interplay between uniform $d$-wave superconductivity and a PDW is precisely what is needed to account for the one-gap, two-gap dichotomy:  On the one hand,
 both orders imply local $d$-wave-like pairing.  On the other hand, globally they produce distinct patterns of broken symmetry, and in particular the PDW has an associated CDW component.  Indeed, as has been noted by several authors \cite{baruch-2008,berg-2008}, the PDW state results in a single-particle spectrum that resembles that seen in the pseudo gap phase immediately above $T_{\rm SC}$,
 with a relatively large gap in the antinodal regions (see Fig.~\ref{fg:fs_gaps}), and a Fermi pocket in the nodal region whose back side (for reasons that were already clear in early DDW calculations \cite{chakravarty08}  of the same quantities) has relatively little spectral weight, giving it the appearance of a Fermi arc.  

As discussed above, the strongest evidence of PDW order comes from transport measurements in roughly 1/8 doped \LBCO, for which (fortunately) ARPES data are available.
 \cite{he09,vall06}.  The ARPES measurements show a clear antinodal gap that changes little on warming into the disordered state; however, below 40 K, there is also $d$-wave-like gap along the nodal arc (although no coherent quasiparticle peaks are seen).  In a weak-coupling analysis, the SDW order that is present would not cause a gap along the near-nodal arc \cite{baruch-2008}, so that one would be led to conclude that the near-nodal gap is due to 
 the existence of a uniform $d$-wave 
component of the order.  This,
 then, creates a problem for the explanation of the frustrated interlayer Josephson coupling.  
However, the SDW order involves substantial Cu moments \cite{luke91,huck08} and so cannot be properly described 
in a weak coupling picture.  Turning to another experimental example, a $d$-wave-like gap (plus a uniform energy offset) has been reported in an ARPES study of Bi$_2$Sr$_{2-y}$La$_y$CuO$_{6+\delta}$ for hole concentrations $x<0.10$ \cite{peng13}.  This regime is insulating \cite{ono03} and a neutron scattering study has found evidence for diagonal spin-stripe correlations for similar dopings in the closely-related Bi$_{2+y}$Sr$_{2-y}$CuO$_{6+\delta}$ \cite{enok13}.  Hence, there is circumstantial evidence that SDW order may cause an (incoherent) $d$-wave-like gap along the near-nodal arc.  
However, further work is obviously needed to resolve these issues. 

\section{Detectable signatures of PDW order}
\label{tests}
To date, the evidence of the existence of a PDW state---even in \LBCO---is indirect.
%,  involving  spectacular transport and optical anomalies that do not have any other obvious explanation. 
  Indeed, even evidence of the existence of  conceptually similar FFLO states in partially magnetized superconductors has been challenging to obtain \cite{mayaffre-2014}.  There are, however, a number of clear experimental signatures which, if observed, would constitute unambiguous evidence of the existence of a PDW.  These signatures  have been discussed previously by us in some detail in \cite{berg-2009b};  here, for completeness, we briefly enumerate some of the most promising such ideas.
\begin{enumerate}
\item[{\bf 1)}] {\bf Existence of a uniform charge $4e$ condensate:} Since in most geometries, the coupling between the PDW order itself and any external superconductor vanishes when spatially averaged, the leading harmonic in the Josephson effects will also vanish.  However, higher order (four-electron) tunnelling processes will always produce phase locking to the accompanying composite charge $4e$ components, $\Delta_{4e,a}$,  of the PDW.  There are several ways this can be detected, for instance in any sort of experiment involving Josephson oscillations in which the relevant Josephson relation is $\hbar \omega = 4$~eV, or if a sliver of \LBCO\ is used as the weak link in a SQUID ring of a conventional superconductor, then the relevant flux quantum should be $hc/4e=\phi_0/2$.
\item[{\bf 2)}] {\bf Tests of interlayer frustration:}  The frustration of the interlayer Josephson coupling apparent in transport and optics in \LBCO\ and other candidate cuprates  is the strongest existing evidence of  a PDW state.  This interpretation can be tested by purposeful perturbations which change the symmetry of the PDW state in such a way as to reduce the interlayer frustration.  As already mentioned, the photo-induced onset of interlayer coherence observed associated with the transient melting of the CDW order may already be an example of such a test.  Other proposed tests involve relieving the interlayer frustration through the application of a suitably oriented in-plane magnetic field \cite{yang-2013}, or possibly by reorienting the stripes by the application of uniaxial strain.
\item[{\bf 3)}] {\bf Half quantum vortices:}  One of the defining features of the PDW is the intertwining of the CDW and SC components of the order parameter.  Since the pairing field must be a single-valued function of position, the phase of the order parameter must be single-valued modulo $2\pi$.  As a consequence, it is easy to show that wherever there is a dislocation in the CDW order, there must be an accompanying half-quantum vortex---a vortex that can be viewed as the fundamental vortex of the uniform charge $4e$ superconducting order.  This is one of the possible topological defects in a PDW---the one with  characteristics that are a unique reflection of the broken symmetry.
\item[{\bf 4)}] {\bf CDW 1${\bf Q}$ order:}  As we have stressed, a composite CDW order with wave-vector ${\bf K}_a = 2{\bf Q}_a$ should be detectable by STM and/or X-ray (or indirectly by neutron) diffraction whenever a PDW with wave-vector ${\bf Q}_a$ is present.\footnote{Modulations in the SC gap can be indirectly observed in STM (in the absence of Josephson tunneling). Evidence of such modulations was given by Fang {\it et al.}\cite{Fang-2004} \cite{baruch-2008}}  However, if below a critical temperature, $T_c$ a uniform SC component develops as well (and if the PDW is not entirely quenched at the same point -- which would only be possible if the transition were first order) then below $T_c$ a {\it subharmonic} component of the CDW order parameter with ordering vector ${\bf K}^\prime_a = {\bf Q}_a$ should develop with an amplitude proportional to the uniform SC order parameter.  
\end{enumerate}

\section{Discussion}
\label{discussion}

Generically, a critical point is reached by varying a single parameter, for instance $T$ in the  theories we have been discussing.  A multicritical point can arise when a second parameter is varied in such a way as to cause two lines of second order transitions to intersect.

{\it Competing orders} generically refers to the physics close to a multi critical point.  In terms of a Landau-Ginzburg-Wilson effective field theory, the multicritical point occurs where the mean-field ordering temperatures, $T_a^0$, of 
two or more orders (labeled by an index, $a$) are equal.  That they ``compete'' implies that the appropriate  biquadratic terms [{\it e.g.}, in Eq.~(\ref{LGW}), $\gamma_j$ with $j=1$--3] are positive.
 Under still more highly fine-tuned circumstances,  a multi critical point can exhibit a higher symmetry.  For instance, 
 the tetra-critical point at which $T_{\rm CDW}=T_{\rm SC}$ in Fig.~\ref{fig:phasediagram1} occurs at a unique temperature and doping concentration, $T_P$ and $x_P$; 
 this tetra-critical point could have a larger $O(4)$ symmetry unifying the CDW and SC orders, but this requires fine tuning  at least one additional parameter.

 {\it Intertwined orders} refers to the case in which
 $T_{\rm SC}^0 \sim T_{\rm CDW}^0$ 
 over a range of situations---{\it i.e.}, in the case of the cuprates, over a range of doping concentrations and material families.  
Where this occurs, it must have its origin in a feature of the microscopic physics as it has no natural explanation simply in terms of robust and generic features of coupled order parameters.    Indeed, any such observation carries with it the suggestion   that  the same features of the microscopic physics that are responsible  for one order also give rise to the other. 
In other words, there likely exists a high energy scale at which an ``amplitude'' of the order develops which cannot really be associated with one or the other order, as it is somehow a precursor to all of them.
Then, at lower scales, small energy effects favor one or the other pattern of ordering.

 In particular, we have in mind the notion that  below $T^*$, to an extent that varies smoothly as a function of $x$, local pairing correlations as well as   local CDW and antiferromagnetic spin correlations begin to grow, in much the way CDW and SC correlations develop below $T^*\sim \Delta_s$ in the quasi-1D model analyzed in Sec.~\ref{phase-diagrams}.   However, as we will now discuss, there are good reasons to question the validity of this perspective.

\subsection{Identification of $T^*$ with local pairing critiqued}
\label{preformedpairs}

That there is a degree of local CDW order over a considerable portion of the pseudo-gap regime of the phase diagram of the cuprates is now an established fact.  Moreover, as discussed extensively in Sec.~\ref{all-together-now} and Sec.~\ref{Pseudogap}, the idea that there are local superconducting pairing correlations which are in one way or another responsible for the $d$-wave character of the pseudo-gap and various subtle precursor indications of superconductivity without long-range phase coherence is an idea that has been advocated in various versions by many authors.  
 Thus, the idea of identifying the pseudogap temperature $T^*$ with the onset of a generalized order parameter amplitude which has both CDW and SC (and possibly AF) character is surely appealing.  In one way or another, it underlies most attempts to come to grips with the phenomenology of intertwined orders.

However, there have been at least as many papers which have critiqued this idea (or 
supposedly ``proven'' it false).  For the most part, these papers have studied a property which might be expected to exhibit a %strong 
signature of local SC (or other) order; then correcting for other contributions (which can, itself, require a rather complicated analysis) they infer from the lack of a clear fluctuational contribution that the conjectured local order does not exist.  For example, constraints on the existence of pairing without phase coherence at temperatures well above $T_{\rm SC}$ have been adduced from measurements of terahertz conductivity \cite{bilb11a}, Nernst effect \cite{wang06}, and magnetoresistance \cite{rull11}.  These constraints are serious---on the other hand,
 as far as we know, there exists no reliable microscopic theory capable of making even semi-quantitative predictions of the consequences of local pairing (or even of sharply defining what this means). 
 Thus, in critiquing this idea, 
 we will 
 instead focus on rather broad phenomenological features of the phase diagram. 

For the most part, the pseudo-gap temperature $T^*$ is defined in one of several somewhat arbitrary ways to extract an explicit number to characterize the scale of $T$ at which the behavior of a particular measured quantity changes from one sort to another.  There thus may be cause to question particular values or even trends in the canonical $T^*$ curves.  
%(Sometimes, $T^*$ can be identified with a sharp change in behavior similar to what is seen at the point of a phase transition or a narrowly rounded phase transition;  in these few cases, $T^*$ can be unambiguously identified.)  
For the purposes of discussion, we will set aside these ambiguities and accept the canonical curves, originally defined such that $T^*(x)$ is the local maximum of the  magnetic susceptibility $\chi$,  as being representative of the pseudo-gap crossover scale.

With this identification, a glaring issue is apparent.  At $x=0$, $T^*$  clearly reflects the local growth of antiferromagnetic correlations; in the 2D AF Heisenberg model, $\chi$ has a maximum at $T^* \sim J/2$ where the AF correlation length is around 2 lattice constants \cite{chakravarty-1988}.  Given that $T^*$ appears to be a continuous function of $x$, for small $x$ this identification must remain valid, implying that $T^*$ has nothing to do with the onset of either SC or CDW correlations.  Conversely,  in many of the hole-doped cuprates for $x \gtrsim 0.1$, while there is still evidence of AF tendencies (local moments) at  short distances, there is an associated spin-gap which grows in magnitude as $T^*$ decreases, making any direct association between $T^*$ and AF order appear unnatural;  on the other hand, it is in the range of doping near $x=1/8$ that the best evidence exists both of short-range CDW order and local SC correlations.   

Much of the trouble may come from a naive expectation that a state with short range order behaves something like a corresponding ordered phase if probed at short distances and times.  Conceptually, near a critical point, the correlations at distances small compared to the correlation length but large compared to the lattice constant look critical---{\it i.e.}, the properties are neither those of the ordered nor of the disordered phase \cite{kivelson-2003}. More specifically, as summarized by example below,  the issue of what is meant by a local magnitude of an order parameter can be much more subtle than any naive intuition would suggest.  However, excuses aside, whether or not local pairing (or singlet formation)  onsets in any well-defined sense at temperatures of order $T^*$ is clearly one of the central unresolved issue in the field.

\subsection{Phase Diagrams of Intertwined Orders}
\label{phase-diagrams}

The physics of strong correlations is largely solved in 1D, and much can be said about the problem in quasi 1D.   
 It is thus worthwhile  summarizing features of this limit.   In Fig.~5, we show the generic phase diagram for an array of weakly coupled LE liquids---a more generic version of the phase diagram already discussed in Fig.~1.  Here, $K_c$ is the charge Luttinger exponent within a single LE chain, which (along with the spin-gap, $\Delta_s$) is an appropriate long-distance measure of the nature of the intra-chain interactions.  There is much here that is reminiscent of the phase diagram of the cuprates.  At high $T$ there is a non-Fermi liquid normal state---in this case it is a set of nearly decoupled LLs.  There is a crossover scale, $T^*\sim \Delta_s$, below which a pseudo-gap opens in the spin fluctuation spectrum.  Finally, at low $T$, there is a complex phase diagram in which two ordered phases compete, and possibly coexist.  In this problem, the ordering temperatures are well separated from $T^*$, as they are proportional to a positive power of the interchain couplings, and so are (by construction) parametrically small.

One aspect of this problem that is worth emphasizing is that both the SC and the CDW susceptibilities are typically small for $T>T^*$, and then begin to grow rapidly for $T<T^*$ (as discussed in Sec.~\ref{LE-ladders}).  Moreover, in this range of $T$, both susceptibilities are proportional to $\Delta_s$, so there is no obvious way to more correctly identify $\Delta_s$ as a precursor pairing or CDW gap.  Rather, the LE liquid is a precursor state in which the opening of the spin gap is conducive to both CDW and SC order.

\begin{figure}[b]
\begin{center}
\includegraphics[width=0.4\textwidth]{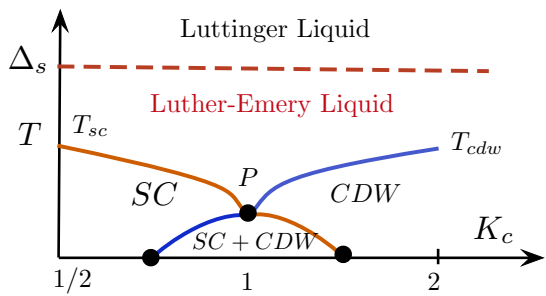}
\caption{(Color online) Sketch of the phase diagram with SC and CDW orders  in  quasi-1D  strongly correlated systems with a spin gap $\Delta_s$. $P$ is a multicritical point.}
\end{center}
\label{fig:phasediagram2}
\end{figure}

Turning from the microscopic to more phenomenological macroscopic considerations, there is a well defined sense in which it is more natural to have multiple orders coexisting in 2D (or quasi-2D) than in higher dimension.  
 For instance, the natural minimal quasi-2D theory of a superconducting
 and an unidirectional incommensurate CDW 
 is obtained by the usual assumption of the Kosterlitz-Thouless theory 
  \cite{chaikin-1995}: At low temperatures the amplitudes of the order parameters 
vary smoothly as a function of parameters, while the phase fields $\theta$  of the SC order parameter, and 
$\phi$  of the sliding  incommensurate CDW order parameter, have strong thermal fluctuations. In this limit the free energy density with  $U(1) \times U(1)$ symmetry is a quadratic functional of the gradients of $\theta$ and $\phi$.   
 Instead of a fixed point, 
  this system has a fixed ``plane'' parametrized by $T/\rho_s$ and $\kappa/\rho_s$. 
   Since gauge invariance forbids any coupling between the two phase fields,  
  vortices and dislocations proliferate independently from each other and their respective Kosterlitz-Thouless  critical temperatures are  $T_{\rm SC}\simeq   (2/\pi) \rho_s$ and $T_{\rm CDW}\simeq   (2/\pi) \kappa$. 

An analysis of this type was done for the thermal melting of the PDW state by \textcite{berg-2009}. 
The general setting was discussed in Section \ref{all-together-now}. To simplify matters, here we will restrict ourselves  to the  case of unidirectional PDW order along the $x$ axis ({\it i.e.}, in a nematic phase). The thermal fluctuations are now
controlled by the effective NLSM in Eq.~\eqref{NLSM-U1s} with the terms corresponding to the $y$ component of the PDW ($\theta_y$ and $\phi_y$) set equal to zero.  There are thus
  three phase fields: 
   $\theta_0$ of the  uniform $d$-wave SC, and 
 $\theta_x$ and $\phi_x$ for the 
SC and CDW components of the 
$x$ directed PDW.  
 The corresponding stiffnesses (ignoring anisotropies) are $\rho_s\equiv K_0 $, $\rho_{\rm PDW}=K_1=K_2$ and $\kappa=K_3=K_4$.  Note that  $V$, when relevant, locks the phase $\theta_0$ to $\theta_x$ (mod $\pi$). 

There are now four types of topological excitations: a) the vortex of the $d$-wave SC phase $\theta_0$, b) the vortex of the PDW phase $\theta_x$, c) 
the half-vortex of  $\theta_x$ bound to a single dislocation of $\phi_x$, and d) the double dislocation of the CDW phase field $\phi_x$. 
Various sequences of vortex unbinding can lead to a variety of complex phase diagrams; an example is sketched in Fig.~6 for the case where $\rho_s$ is assumed to be small compared to $\rho_{\rm PDW}$ and $\kappa$.   The considerations (which are described in more detail in \textcite{berg-2009}) which lead to this phase diagram are as follows:
 a)  At low temperature, there is a fully ordered ``striped SC,'' in which $\theta_0$ and $\theta_{x}$ are locked to each other, and there is coexisting CDW order.
 b) By assumption, the uniform SC order melts above a low temperature, $T_d$, where the $\theta_0$ vortices  unbind through a KT transition.  Above $T_d$, the large fluctuations of $\theta_0$ render $V$ irrelevant as well.  The resulting phase is a pure PDW.
 c)  If $\kappa/\rho_{\rm PDW}$ is large, the next transition involves the proliferation of $\theta_{x}$ vortices, leading to a KT transition to a pure CDW phase.  Then, at still higher temperatures, the proliferation of $\phi_x$ dislocations leads to a uniform non-superconducting nematic phase.
 d) If  $\kappa/\rho_{\rm PDW}$ is small, the first transition from the PDW phase involves the proliferation of $\phi_{x}$ double dislocations, resulting in  a KT transition to a uniform superconducting phase -- one, however, with the charge $4e$ condensate from Eq.~(\ref{4e}).  Then, at  higher temperatures, the proliferation of $\theta_x$ half-vortices (reflecting the charge of the condensate) leads again to the uniform non-superconducting nematic phase.
 e) At intermediate values of $\kappa/\rho_{\rm PDW}$, the lowest energy topological excitations are the bound state of the half-vortex and a single dislocation, the proliferation of which results in a direct KT transition to the pure nematic phase.  In all cases, at still higher temperatures, the proliferation of disclinations (not included in the NLSM we have discussed) will lead to an Ising transition from the nematic phase to a uniform ``normal'' phase, as indicated by $T_n$ in Fig. 6.

 Some details of this pattern may be altered  significantly by changing parameters [{\it i.e.}, the stiffnesses but also the strength of the coupling to the lattice \cite{barci-2011}]. The phase diagram of Fig.~6 has several multicritical points (denoted by $P$ and $P'$) which have a larger ``emergent'' [$SU(3)$!] symmetry \cite{berg-2009}.
It is important 
to note that  
in this problem,
 {\em  the natural $U(1)$ symmetries of the order parameters alone 
  produce a phase diagram  with a complex set of phases, and critical temperatures  that are 
  comparable  
  to each other} without 
  invoking any fine-tuned multicritical point with a large emergent symmetry. 

\begin{figure}[t]
\begin{center}
\includegraphics[width=0.4\textwidth]{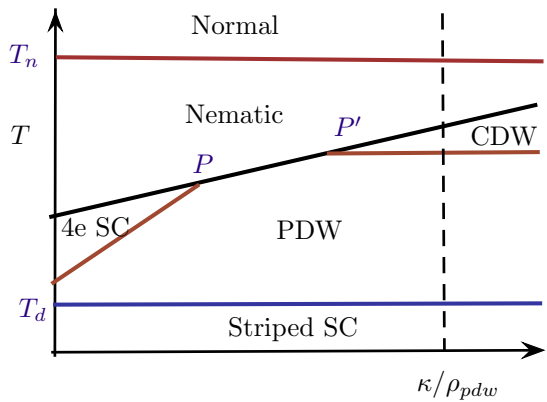}
\caption{(Color online) Qualitative phase diagram for the melting of a unidirectional PDW state coexisting with $d$-wave SC order. See text for details. {\LBCO} is presumably along dark broken vertical line.}
\end{center}
\label{phasediagram3}
\end{figure}

Finally, it is important to mention that evidence, both theoretical and experimental, has been adduced concerning the existence of other new phases of matter in the cuprates including DDW order  \cite{chakravarty-2001c}, intra-unit cell orbital current order (OAF)  \cite{varma-2006b}, and various forms of topologically ordered phases.  It goes without saying that establishing which of these phases actually exists in this family of materials is of central importance. From the present perspective, the existence of any of these phases would primarily serve to further emphasize the intrinsic complexity of the phase diagram.  While some of these orders, for instance some forms of OAF order, can be constructed as composite orders in terms of the fields already present in our analysis, others (such as DDW) would involve the introduction of yet further fields.

\subsection{Critique of theories with emergent symmetries} 
\label{subsec-critique}

Theories invoking large emergent symmetry groups have been proposed in the context of high $T_{\rm SC}$ superconductors, starting with S. C. Zhang's proposal to unify $d$-wave superconductivity (which has a complex order parameter field) with N{\'e}el antiferromagnetic order (which has a three-component real order parameter field) in a larger $SO(5)$ symmetry \cite{zhang-1997,demler-2004}. Generalizations of this concept have since been pursued, including a model  relating nematic order to $d$-density wave and $d$-wave superconductivity (where the larger symmetry is $SO(6)$) by Kee and coworkers \cite{kee-2008,kee-2010},
and, most germane to the present discussion, theories of  
 \textcite{sachdev-2013},  \textcite{hayward-2014}, and  \textcite{Efetov-2013}
which envisage an $SO(6)$ or $SU(2)$ symmetry relating charge-density-wave order and superconductivity.

What is very attractive about  these approaches is that they intertwine the various orders to such an extent that they become indistinguishable at short distances. However, typically, systems are less symmetric at low energies and long wave-lengths than at the microscopic level---this observation underlies, for example,  the Standard Model of particle physics that unifies the electromagnetic, weak, and strong interactions at high energies.  Nevertheless, emergent symmetries are not unheard of.  Both the Kondo impurity and the two-channel Kondo impurity problems exhibit emergent $SU(2)$ spin-rotational symmetry at low energies, even if this symmetry is broken strongly in the microscopic model.  The two-leg Hubbard ladder 
exhibits a fully gapped phase which, for small $U$ (where all the gaps are exponentially small), has an emergent $SO(8)$ symmetry \cite{lin-1998}. \footnote{Intertwined orders in $d=2$ (with possibly enhanced symmetries) have been found in weakly correlated systems with band-structures with quadratic crossings \cite{sun-2009,murray-2014,vafek-2014}.}
Several authors \cite{fernandes-2010,sachdev-2013,
 Efetov-2013,davis-2013} have suggested
that an emergent symmetry unifying CDW and SC order can arise 
from singular induced interactions 
 at ``hot-spots'' on the Fermi surface, 
such as occur in  close proximity to an antiferromagnetic QCP of a Fermi liquid.

The behavior of such multicritical points with enhanced symmetries is well understood and has been studied since the early 1970's
 \cite{aharony-1974,nelson-1975,kosterlitz-1976,aharony-2003}. The main results of these renormalization group studies is that, except for a special case which will not concern us here, the enhanced symmetry is fragile, since at the fixed point of   
such multicritical points 
various symmetry breaking operators 
  are  relevant. In particular, at an $SO(4)$ invariant multicritical point, terms that break the symmetry to  $O(2) \times O(2)$ are strongly relevant.

Calabrese and coworkers have further investigated the stability of the multicritical points associated with breaking of a larger symmetry group to smaller subgroups, $O(n_1+n_2) \mapsto O(n_1) \times O(n_2)$, using a five-loop $\epsilon$ expansion \cite{calabrese-2003}.
 They find that for $N=n_1+n_2\geq 3$, the symmetry-breaking perturbations render the $O(N)$ symmetric fixed point  unstable and that, for $N\geq 5$, RG flows drive the system to a critical decoupled fixed point at which the  $O(n_1)$ and $O(n_2)$ symmetries are decoupled. For $N=3$ (and possibly also for $N=4$) the actual critical behavior is controlled by  the so-called   biconical fixed point. Furthermore the crossover away from the unstable $O(N)$ fixed point is quite rapid reflecting the fact that the symmetry-breaking perturbations are strongly relevant. In particular (and this is what matters to our analysis), near the multicritical point the actual resulting $T_c$'s rapidly diverge from each other  as a power law of the form $|T_{\rm SC}-T_{\rm CDW}|\propto |r_{\rm SC}-r_{\rm CDW}|^{1/\phi}$. The crossover exponent $\phi$ is determined by the scaling law $\phi=\nu (d-\Delta_2)$ where $\nu$ is the correlation length exponent at the $O(N)$-symmetric fixed point, $d$ is the dimension of space and $\Delta_2$ is the scaling dimension 
of the quadratic symmetry breaking operator $|\Delta(\bm x)|^2-|\rho_{\bm Q}(\bm x)|^2$. Current best estimates \cite{calabrese-2003} for the crossover exponent in $d=3$ dimensions yield $\phi\simeq 1.35$ for $N=4$ (and somewhat larger values for $N\geq 5$) which implies that the splitting of the critical temperatures is {\em larger} than a linear function of the quadratic symmetry-breaking field. In addition, a further complication is that, over a significant range of parameters, these phase transitions have a strong tendency to become fluctuation-induced first order transitions. 

Many of these issues were discussed extensively in the context of the $SO(5)$ theory [see, {\it e.g.}, \textcite{aharony-2003}]. The upshot of this analysis is that, even if the transition is actually continuous, the fixed point with high symmetry is generally unstable and that for the critical temperatures of the competing orders to be similar in magnitude requires 
extremely delicate 
fine-tuning of the microscopically-determined control parameters. This is the key message from this analysis.

The high temperature superconductors are quasi-two-dimensional systems. However, 
the fine-tuning problem, if anything, 
 is worse in 2D (and quasi-2D) systems with a large 
 continuous global symmetry, 
  such as $O(N)$ with $N>2$.  
 In strictly 2D, such systems cannot have a phase transition at any finite temperature. Close to $d=2$  they are well described by a non-linear $\sigma$-model, which describes the long-distance fluctuations of a {\em classical} $N$-component Heisenberg model, whose order parameter is an $N$-component unit vector, ${\bm n}(\bm x)$ (such that $|| \bm n ||^2=1$). 
Renormalization group analysis of the two-dimensional  non-linear $\sigma$-model \cite{polyakov-1975,brezin-1976} shows that thermal fluctuations are marginally-relevant (``asymptotically free'') perturbations at the $T=0$ fixed point, and consequently  that 
this system is in its disordered (high temperature) 
phase for all values of the temperature $T$. As a result, at low temperatures  the correlation length  scales as $\xi(T) \sim a \exp[2\pi K/(N-2)T]$, where $K$ is the helicity modulus and $a$ is the lattice spacing. 

 This gives a broad fluctuational regime which
 is {\it exponentially sensitive} to the magnitude of coupling parameters of the symmetry-breaking operators and/or to three-dimensional couplings. 
 An important example is the case in which there is a small symmetry breaking term of (dimensionless) magnitude $h$ which explicitly breaks an  $O(N)$ symmetry (with $N>2$)  down to $ O(2)\cong U(1)$. This problem was discussed extensively by \textcite{affleck-1986} (in the context of easy-plane symmetry-breaking in quantum antiferromagnetic spin chains) and, more recently, by \textcite{fellows-2012} in the context of competing orders. A key consequence of the marginal relevance of temperature in the $O(N)$ symmetric theory is that the (Kosterlitz-Thouless) critical temperature of the $O(2)$-invariant system has a logarithmic dependence on the 
 symmetry breaking field $h$
 \cite{affleck-1986,fellows-2012}
 \begin{equation}
 T_{KT} \sim \frac{K(N-2)}{4\pi} \frac{1}{\ln(1/|h|)}
 \end{equation}
 Since the critical temperature of the $O(N)$-symmetric model is zero, the finite value of $T_{KT}$ 
reflects the significance of even very small symmetry breaking terms.

Finally, we discuss how these considerations apply to the multiple orders considered by 
 \textcite{hayward-2014} in the context of the cuprates, as representative of the various proposals for emergent higher symmetries in the cuprates introduced at the beginning of this section.  The underlying problem has a $U(1)\times U(1)\times U(1)\times Z_2$ symmetry, where the first $U(1)$ is associated with the superconducting order, and the final two $U(1)$'s and the $Z_2$ are associated with the CDW order, and correspond to translational symmetry in the $x$ and $y$ directions and rotation by $\pi/2$ about the $z$ axis, respectively.  Hayward {\it et al.}\ assume that 
 there is an approximate much larger $SO(6)$, although they do explicitly take into account terms that differentiate the superconducting and CDW components of the order parameter, which thus break the symmetry down to $U(1)\times SO(4)$. However, the  assumed $SO(4)$ symmetry is still non-generic, and this,  along with the assumption that interlayer coupling can be neglected, allowing the system to be treated as 2D, is what is responsible for the central feature of the scenario of \textcite{hayward-2014}, resulting in a  CDW correlation length that never diverges at any non-zero temperature.  Again, terms that violate either assumption lead to a CDW ordering temperature that is only small in proportion to the inverse of the logarithm of the term's magnitude.

As a possible rejoinder to this critique, several studies \cite{Efetov-2013,sachdev-2013,meier-2014}  have proposed that a near-perfect  symmetry between CDW and SC orders can result from a higher level organization associated with close proximity to a metallic quantum critical point associated with  large $Q$  antiferromagnetic order.  These theories consider a  metallic Fermi liquid which is not too strongly coupled to the quantum critical fluctuations, so that they  have a strong effect only on the
electronic quasiparticles residing very close to a set of ``hot-spots'' on the Fermi surface \cite{abanov-2000,vekhter-2004,chubukov-2005,tsvelik-2014} (for a review see \textcite{wang-2014b}).
It is unclear whether the weak-coupling focus on hot-spots is reasonable in realistic, strongly coupled systems.
  Indeed, quantum Monte-Carlo studies of  \textcite{Berg-2012} of a metallic antiferromagnetic quantum critical point found clear evidence of induced superconductivity, but no reported evidence of the growth of CDW correlations of comparable strength.  Nevertheless, even if induced CDW order  can also arise as a consequence of scattering by quantum critical AF fluctuations, it is far from clear why the proposed emergent symmetry should, at finite temperatures especially, be immune to the already discussed divergent flows of the symmetry breaking terms associated with classical critical phenomena.  Moreover, for this scenario to apply, the system must at least be fine-tuned to the near proximity of an antiferromagnetic quantum critical point.  In many of the hole-doped cuprates in much of the range of dopings where SC and CDW orders appear to be intertwined, the antiferromagnetic correlation length measured in neutron scattering is no more than 1-2 lattice constants\footnote{The equal-time spin correlation length, $\xi$, is obtained by integrating the magnetic $S({\bf Q},\omega)$ over $\omega$, determining the half-width-at-half-maximum of the resulting $S({\bf Q})$, and taking the inverse of that width.  For La$_{1.86}$Sr$_{0.14}$CuO$_4$, such an analysis gave $\xi\approx a$ \cite{hayd96a}.  For optimally-doped \YBCO\ and \BSCCO, one can estimate an upper limit from the half-width of the resonance peak in the superconducting state, yielding $\xi \lesssim 1.3a$ \cite{dai01} and $\xi\lesssim a$ \cite{fong99}, respectively.}---{\it i.e.}, nowhere near being quantum critical.
 
\subsection{Ineluctable Complexity}
\label{ineluctable}

The intrinsic complexity of the phase diagrams of  correlated materials---the cuprates in particular---is by now self evident.  At the  microscopic level, this is clearly a result of the 
quantum frustration of such systems:  the kinetic energy favors highly delocalized uniform density fluid states, while the interaction energy favors localized, spatially inhomogeneous crystalline states.  Understanding each phase that occurs with great precision is certainly a worthwhile undertaking. What is  presented here is a step toward understanding the origin of the complexity itself.  The intertwining of CDW, SDW, and SC order in the cuprates is particularly striking and well documented, and  is probably \cite{emery-1993,Fradkin-2012,dagotto-2005} associated with a rather general local tendency to phase separation.  

We have presented highly suggestive, but by no means conclusive, evidence for the existence of a PDW phase in the cuprates.   If confirmed, this represents the discovery of a new phase of matter, which would be  significant independent of any other implications.  More broadly, as a state that tangibly intertwines CDW and SC (and, in some cases, SDW order as well) in an explicit, testable fashion ({\it i.e.}, associated with a pattern of broken symmetry),  it has the potential for providing a unifying starting point for studying the broader issues of intertwined orders in the cuprates.

\begin{acknowledgements}
We thank Erez Berg,  Vadim Oganesyan,  ZX Shen,  Eun-Ah Kim, Akbar Jaefari, Daniel Barci, Aharon Kapitulnik, Seamus Davis, Gabe Aeppli, Subir Sachdev, Patrick Lee, Hae-Young Kee, Sudip Chakravarty, Dung-Hai Lee, Laimei Nie, Akash Maharaj, Sam Lederer, Rodrigo Soto Garrido, Marc-Henri Julien, Ian Fisher, Andrea Damascelli, Ali Yazdani, Louis Taillefer, and Ruihua He for stimulating discussions and suggestions.  EF and SAK thank the Kavli Institute for Theoretical Physics (and the Simons Foundation) and the KITP IRONIC14 program for support and hospitality. This work was supported in part by the National Science Foundation through grants NSF DMR-1064319 and NSF DMR-1408713 at the University of Illinois (EF), NSF DMR-1265593 (SAK), PHY11-25915 at KITP (EF,SAK),  and by the U.S. Department of Energy (DOE), Office of Basic Energy Sciences (BES), Division of Materials Sciences and Engineering (MSE) under Award No.\ DE-SC0012368 through the Materials Research Laboratory of the University of Illinois (EF)  and DE-AC02-76SF00515 at Stanford (SAK).  JMT is supported at Brookhaven by the U.S. DOE, BES, MSE, through Contract No.\ DE-AC02-98CH10886.
\end{acknowledgements}

%merlin.mbs apsrmp4-1.bst 2010-07-25 4.21a (PWD, AO, DPC) hacked
%Control: key (0)
%Control: author (75) reversed first initials jnrlst
%Control: editor formatted (0) differently from author
%Control: production of article title (-1) disabled
%Control: page (0) single
%Control: year (1) truncated
%Control: production of eprint (0) enabled
%

%\bibliographystyle{apsrmp4-1}
%\bibliography{htsc,LNO}

\end{document}